%% file: main.tex
\documentclass[10pt,letterpaper]{article}
\usepackage[margin=0.75in,top=0.72in,bottom=0.75in]{geometry}
\usepackage[utf8]{inputenc}
\usepackage{times}
\usepackage{microtype}
\usepackage{amsmath}
\usepackage{amssymb}
\usepackage{graphicx}
\usepackage{booktabs}
\usepackage[font=small,labelfont=bf]{caption}
\usepackage{array}
\usepackage{tabularx}
\usepackage{multirow}
\usepackage{xcolor}
\definecolor{alethiacite}{HTML}{001472}
\usepackage{textcomp}
\usepackage[round,authoryear]{natbib}
\let\cite\citep
\usepackage{url}
\usepackage{xurl}
\usepackage[colorlinks=true,linkcolor=black,citecolor=alethiacite,urlcolor=alethiacite]{hyperref}
\usepackage{flafter}
\usepackage{placeins}
\usepackage{stfloats}
\usepackage{float}
\input{metadata}

\makeatletter
\renewcommand\section{\@startsection{section}{1}{\z@}%
  {-1.5ex \@plus -0.5ex \@minus -.2ex}%
  {0.8ex \@plus .2ex}%
  {\raggedright\normalfont\large\bfseries}}
\renewcommand\subsection{\@startsection{subsection}{2}{\z@}%
  {-1.25ex\@plus -0.4ex \@minus -.2ex}%
  {0.55ex \@plus .15ex}%
  {\raggedright\normalfont\normalsize\bfseries}}
\renewcommand\subsubsection{\@startsection{subsubsection}{3}{\z@}%
  {-1ex\@plus -0.35ex \@minus -.2ex}%
  {0.4ex \@plus .1ex}%
  {\raggedright\normalfont\normalsize\itshape}}
\makeatother
\renewenvironment{abstract}{\section*{\centering Abstract}\small}{\par}

\hypersetup{
  pdftitle={\papertitle},
  pdfauthor={\paperauthor},
  pdfsubject={Benchmarking Neural Defend ARCAS 290M for AI-generated audio detection},
  pdfkeywords={AI-generated audio detection, foundation model, audio deepfake detection, speech anti-spoofing, robustness, multilingual, partial fakes, replay, benchmark evaluation}
}
\renewcommand{\topfraction}{0.95}
\renewcommand{\bottomfraction}{0.85}
\renewcommand{\textfraction}{0.03}
\renewcommand{\floatpagefraction}{0.65}

\newcommand{\NA}{\textemdash}

\begin{document}
\twocolumn[
\begin{center}
\vspace*{0.35in}
\rule{\textwidth}{0.8pt}\par
\vspace{0.32in}
{\LARGE\bfseries \papertitle\par}
\vspace{0.28in}
\rule{\textwidth}{0.8pt}\par
\vspace{0.32in}
{\normalsize\bfseries Sivashankar Selvarajan\textsuperscript{1}\quad Piyush Verma\textsuperscript{1}\quad Sumit Kumar\textsuperscript{1}\quad Sharayu N. Deshmukh\textsuperscript{1}\par}
\end{center}
\vskip 0.28in
]
\footnotetext[1]{Neural Defend Inc. Correspondence to: Sivashankar Selvarajan \texttt{<siva@neuraldefend.com>}.}

\begin{abstract}
Voice cloning now needs only seconds of audio, and detectors tuned to one dataset often fail when the generator, channel, or language changes. We benchmark ARCAS 290M, Neural Defend's foundation model for detecting AI-generated audio. A single frozen production model, with no benchmark-specific fine-tuning, generalizes across text-to-speech, voice-conversion, and codec-language-model generators, telephone and Bluetooth channels, recording conditions, and languages. We evaluate it on the complete official test sets of ten public benchmarks, scoring every recording exactly as released, and against commercial systems on independent leaderboards. ARCAS 290M ranks first among all systems on the Podonos leaderboard and has the lowest equal error rate among commercial detectors on ASVspoof~5. It separates real from AI-generated speech almost perfectly on ASVspoof~5, SpeechFake, and ShiftySpeech, and detects nearly all AI-generated speech in Hindi, Tamil, Bengali, and Telugu. On every benchmark with a published zero-shot comparison, including in-the-wild audio, partial fakes, and the conditions hardest for the field, it surpasses the strongest reported detector by a wide margin. These results establish ARCAS 290M as a general-purpose foundation for AI-generated audio detection in fraud prevention, platform safety, and media verification.
\end{abstract}

\input{sections/01_introduction}
\input{sections/02_related_work}
\input{sections/03_system_scope}
\input{sections/05_benchmarks}
\input{sections/07_discussion}

\bibliographystyle{plainnat}
\bibliography{references}

\clearpage
\onecolumn
\input{sections/09_appendices}

\end{document}

%% file: metadata.tex
\newcommand{\papertitle}{Benchmarking Neural Defend ARCAS 290M: A Foundation Model for AI-Generated Audio Detection}
\newcommand{\paperauthor}{Sivashankar Selvarajan, Piyush Verma, Sumit Kumar, and Sharayu N. Deshmukh}

%% file: sections/01_introduction.tex
\section{Introduction}

Generative speech models can now clone a voice from a few seconds of reference audio, and human listeners identify current synthetic speech as fake only about 69\% of the time \cite{muller2026eroding}. AI-generated audio is already used for voice phishing, impersonation of executives and public figures, and fabricated evidence. Reliable detection has therefore become core infrastructure for call centers, banks, platforms, and newsrooms, and it must hold up on audio that no laboratory dataset anticipated.

Generators change every few months; audio reaches the detector through codecs, telephone networks, loudspeakers, and social platforms; and an attacker may alter a single word rather than a whole recording. Published detectors are typically trained and tested within one benchmark and lose much of their accuracy when the generator, channel, or language changes \cite{speechfake2025,mlitw2026,deepfakeeval2024}. A production system must instead behave as a foundation model for the task: one model that transfers, without retraining, across generators, channels, languages, and recording conditions.

In this report we benchmark ARCAS 290M, Neural Defend's foundation model for AI-generated audio detection. We score the complete official test sets of ten public benchmarks, 5,644,079 recordings in total, exactly as released: ASVspoof~5, SpeechFake, ShiftySpeech, CodecFake+, ML-ITW, EchoFake, HQ-MPSD, LlamaPartialSpoof, IndicSynth, and PartialEdit. They span current text-to-speech, voice-conversion, and codec-language-model generators; codec, telephone, and Bluetooth channels \cite{wang2024asvspoof5}; partial manipulations \cite{llamapartialspoof2024,hqmpsd2025,partialedit2025}; physical replay \cite{echofake2025}; social-media audio \cite{mlitw2026}; and more than 15 languages, including four Indian languages \cite{indicsynth2025}. The model is evaluated as deployed, through its production endpoint with no benchmark-specific fine-tuning, following the protocol of the companion ARCAS 1B evaluation \cite{selvarajan2026arcas1b}.

ARCAS 290M ranks first of 28 systems on the independent Podonos leaderboard, with 99.93\% accuracy \cite{podonos2026benchmark,arcas2026podonos}, and has the lowest ASVspoof~5 EER (0.14\%) among the commercial detectors reported on that set \cite{dowerah2026dfarena,zhu2024learnfromreal}. Across the five benchmarks of whole-file AI-generated speech it detects 99.60\% of 4.80 million fakes. On every benchmark with a published zero-shot comparison, its EER is less than half that of the strongest reported detector, and more than ten times lower on codec-language-model speech and on HQ-MPSD partial fakes.

This evidence differs from how detection performance is usually reported. Vendors typically cite accuracy on internal test sets whose composition is not released, and academic papers often subsample or re-encode test sets. Here every number comes from a complete, unmodified public test set or an independent private-label leaderboard, with exact denominators and confidence intervals, so each claim can be checked against the released data.

Our contributions are:
\begin{itemize}
\item \textbf{A foundation model for AI-generated audio detection.} A single production model that detects fully and partially AI-generated speech across generators, codecs, channels, and languages, with no task-specific fine-tuning.
\item \textbf{A head-to-head comparison with commercial systems.} Results on the same test data as commercial detectors, on the Podonos leaderboard and the Speech DF Arena, so buyers can compare vendors directly.
\item \textbf{Full-scale, condition-level evaluation.} Every recording of ten official test sets, broken down by generator, codec and channel, replay setup, language, platform, and fake-segment share, including the conditions hardest for all detectors and where human review adds most value.
\end{itemize}

%% file: sections/02_related_work.tex
\section{Related Work}

\subsection{Generalization in synthetic-speech detection}

The ASVspoof challenges established the protocol of training on one set of attacks and testing on unseen ones. ASVspoof~5 extends it to crowdsourced Multilingual LibriSpeech speech, 16 unseen evaluation attacks including adversarial filters, and 11 codec and telephone conditions \cite{wang2024asvspoof5,wang2025asvspoof5db,asvspoof5summary2026}. Its best open-condition system reached 2.59\% EER on the Track~1 evaluation set, and its AASIST baseline 29.12\% \cite{wang2024asvspoof5}.

Later datasets show that generalization fails along several axes. SpeechFake and MLAAD broaden generator and language coverage to tens of text-to-speech (TTS) and voice-conversion (VC) systems \cite{speechfake2025,mlaad2024}. CodecFake+ targets speech produced by neural-codec language models and real speech re-synthesized through neural codecs \cite{codecfakeplus2025}. ShiftySpeech holds the vocoder fixed and shifts the source domain, language, and speaker \cite{shiftyspeech2025}. Across these studies, detectors trained on ASVspoof 2019 reach EERs of roughly 19--39\% on the newer sets \cite{speechfake2025,codecfakeplus2025}. Generalization, robustness to channel effects, calibration, and operating-point behavior are distinct properties, and we report them separately.

Baseline detectors fall into a few families. Spectral and raw-waveform models such as LFCC-based classifiers, RawNet2, and AASIST learn artifacts directly from the training attacks. Self-supervised front-ends such as wav2vec~2.0 and XLS-R transfer better because they are pre-trained on large amounts of real speech \cite{speechfake2025,codecfakeplus2025}. Their gains are largest in-domain: a W2V+AASIST model trained on SpeechFake's bilingual split reaches 3.54\% EER on its own test set but 23.78\% when trained on ASVspoof 2019 instead \cite{speechfake2025}. Training-data diversity matters as much as architecture. On ML-ITW, the same XLSR+AASIST model ranges from 49.40\% EER when trained on ASVspoof 2019 to 27.65\% when trained on SpoofCeleb \cite{mlitw2026}.

\subsection{Benchmarks and performance interpretation}

A benchmark number depends on the test population as much as on the detector. Accuracy changes with the real-to-fake ratio, which in our test sets ranges from about 1:1 (EchoFake) to about 1:12 (LlamaPartialSpoof). Subsampled or re-balanced test sets make cross-paper comparison unreliable. We therefore report EER and class-wise rates alongside accuracy and record exact denominators.

Recent benchmarks target conditions under which detectors fail. LlamaPartialSpoof splices LLM-rewritten phrases into real recordings \cite{llamapartialspoof2024}; HQ-MPSD builds artifact-controlled partial fakes in eight languages with added noise and echo \cite{hqmpsd2025}; and PartialEdit changes a single word with neural speech editors \cite{partialedit2025}. Detectors trained on PartialSpoof reach 24.49\%, 51.38\%, and 23.72\% EER on these three sets, respectively \cite{llamapartialspoof2024,hqmpsd2025,partialedit2025}.

EchoFake replays real and fake speech through loudspeakers and re-records it; its zero-shot baselines score 37.81--56.23\% EER on the open set \cite{echofake2025}. ML-ITW collects social-media deepfakes in 14 languages, on which eleven academic detectors score 40.49--50.03\% EER \cite{mlitw2026}. IndicSynth reports 8--93\% EER for ASVspoof-trained models across Indian languages \cite{indicsynth2025}.

Deepfake-Eval-2024 is one of the few open evaluations of commercial detectors: on audio circulated in 2024, open-source models lost about 48\% of their AUROC, while the best of eight commercial audio detectors reached 0.89 accuracy and 0.93 AUROC \cite{deepfakeeval2024}. The Speech DF Arena evaluates open and proprietary detectors on 14 datasets, including ASVspoof~5 \cite{dowerah2026dfarena}, and the Podonos leaderboard scores commercial APIs and open models on a private-label set of modern voice-cloning audio \cite{podonos2026benchmark}. On both, proprietary systems far outperform open models trained on ASVspoof 2019.

Metric choice is a second source of divergence. The ASVspoof~5 challenge ranks systems by minimum detection cost (minDCF), reports actual cost (actDCF) and $C_\mathrm{llr}$ to expose calibration, and keeps EER as a secondary threshold-free measure \cite{wang2024asvspoof5}. Most other benchmarks report EER only, and partial-fake studies add segment-level EER \cite{partialedit2025,robustloc2025}. A detector can therefore look strong under one metric and weak under another. We report each benchmark in its native population and treat cross-paper numbers as context rather than as a controlled ranking.

\subsection{Audio-only evaluation}

In deployment, a detector score is one signal within a wider authenticity review that may include caller metadata, video, provenance records, and human judgment. An audio-only benchmark measures how well that signal separates real from synthetic speech under stated conditions. It does not measure the accuracy of a full review process, attribution of a fake to a specific generator, or resistance to an attacker who adapts to the detector.

%% file: sections/03_system_scope.tex
\section{Methodology and Evaluation Design}

The evaluation follows one principle: each benchmark is scored as its authors released it. We download the complete official test set, send every file unchanged to the production endpoint, and compute all metrics from the returned verdicts and scores. No record is sampled, converted, or added from another corpus, and no benchmark-specific adjustment is made. This section describes the evaluated system, how records are accounted for, how decisions and metrics are derived, and how uncertainty is reported.

\subsection{System Scope and Reproducibility}

ARCAS 290M (version v3.1d) is Neural Defend's production foundation model for AI-generated audio detection. All records of all ten benchmarks returned a valid response; no record is unscored.

The detector is accessed through its HTTP inference endpoint, which accepts one audio file per request and returns a verdict and a confidence $c_i$. For an audio record $x_i$, we derive a fake score
\begin{equation}
s_i=\begin{cases} c_i & \text{if the verdict is \emph{AI-Generated}},\\ 1-c_i & \text{if the verdict is \emph{Genuine}}.\end{cases}
\label{eq:score}
\end{equation}
One frozen detector and decision policy were used throughout, with no fine-tuning, calibration, or benchmark-specific update. Because the endpoint reports no model version, 20 fixed canary records were re-scored every 30 minutes throughout the run; all returned bit-identical outputs. Re-scoring clips of an earlier 2,000-record pilot reproduced every earlier output exactly.

\subsection{Evaluation Design and Statistical Methods}

\subsubsection{Study Design and Record Accounting}

\paragraph{Population and labels.}
The study includes ten public benchmarks (Table~\ref{tab:coverage}). For each, we use its complete official test or evaluation release, with labels from the authors' protocol files. Where no test split is published and the dataset is released for evaluation, we use the full release. Files are uploaded exactly as distributed (FLAC, WAV, MP3, or audio bytes stored in Parquet) at their original sample rate. Records use $y_i=0$ for real speech and $y_i=1$ for fake speech. Denominators use scored records.

PartialEdit and IndicSynth are released without real speech. For these we report recall on fakes only. We do not pair them with real audio from other corpora, because recording and format differences would confound the comparison.

Coverage accounting is kept separate from detector error accounting. A record is scored if the endpoint returns a valid verdict. HTTP errors, undecodable files, and files larger than 50~MiB are recorded as unscored and reported as coverage, never as correct or incorrect. Every file sent to the endpoint is byte-identical to the released file; this was verified by SHA-256 on a sample of files.

\paragraph{Decision rule and confusion counts.}
The predicted label $\hat y_i$ is the endpoint's own verdict; we apply no threshold of our own. This is equivalent to $\hat y_i=\mathbb{1}[s_i>0.5]$ except at $s_i=0.5$ (a confidence of exactly 0.5), where the endpoint returns either verdict; in that case the returned verdict is used. Scores are rounded to four decimals, so threshold-free curves step through each distinct score value. With fake speech as the positive class,
\begin{align}
N&=TP+TN+FP+FN,\\
\mathrm{Acc}&=\frac{TP+TN}{N},\\
\mathrm{TPR}&=\frac{TP}{TP+FN},\quad
\mathrm{TNR}=\frac{TN}{TN+FP},\\
\mathrm{Precision}&=\frac{TP}{TP+FP},\\
F_1&=\frac{2TP}{2TP+FP+FN}.
\end{align}
Where presentation-attack terminology is useful, fake speech is the attack class, so $\mathrm{APCER}=1-\mathrm{TPR}$ and $\mathrm{BPCER}=1-\mathrm{TNR}$.

\paragraph{Partial fakes and utterance-level labels.}
The endpoint returns one verdict per file, so every benchmark is evaluated at utterance level. A record that contains any synthetic or edited segment is labeled fake, following each dataset's protocol. Segment-level localization metrics are therefore not computed. Fake-segment length, where released, is used only to group records.

\paragraph{Failure slices and baseline comparison.}
Failure slices are defined only from metadata released with each benchmark, such as generator, codec or channel, replay device, language, and platform; we do not infer or add labels. Published baselines are quoted from each dataset's own paper and placed next to our results only when they use the same test population. Each baseline is marked as in-domain or zero-shot.

\subsubsection{Metrics and Aggregation}

\paragraph{Ranking metrics and native metrics.}
The EER and AUROC are computed from $s_i$ and are independent of the decision rule. For ASVspoof~5, we run the official Track~1 evaluation package (minDCF, actDCF, $C_\mathrm{llr}$) \cite{wang2024asvspoof5}. The endpoint returns a capped probability rather than a calibrated log-likelihood ratio, so actDCF and $C_\mathrm{llr}$ describe the deployed system rather than its best operating point.

\paragraph{Aggregation and uncertainty.}
All metrics are pooled at record level within each benchmark. Proportions carry 95\% Wilson score intervals,
\begin{equation}
\frac{\hat p+\frac{z^2}{2n}\pm z\sqrt{\frac{\hat p(1-\hat p)}{n}+\frac{z^2}{4n^2}}}{1+\frac{z^2}{n}},\qquad z=1.96,
\end{equation}
and EER and AUROC intervals use 1,000 bootstrap resamples drawn separately within the real and fake classes, so each resample keeps the original class counts (fixed seed 12345). Slices with fewer than 30 real or 30 fake records are reported but marked as too small to judge.

%% file: sections/05_benchmarks.tex
\section{Results}

The ten benchmarks are ordered below by decreasing pooled accuracy in Table~\ref{tab:main_metrics}: ASVspoof~5, SpeechFake, ShiftySpeech, CodecFake+, ML-ITW, EchoFake, HQ-MPSD, and LlamaPartialSpoof, then IndicSynth and PartialEdit, which have no real speech, by recall on fakes. Populations and native metrics differ, so this order is a reading aid rather than a cross-benchmark leaderboard. Among these benchmarks, third-party results for commercial detectors exist only for ASVspoof~5, through the Speech DF Arena \cite{dowerah2026dfarena} and Reality Defender's published challenge system \cite{zhu2024learnfromreal}; for the other nine, ARCAS 290M is compared with the baselines published by each dataset's authors. Results on the independent Podonos leaderboard of commercial detectors, which scores current voice-cloning audio against private labels, are reported in a companion report \cite{arcas2026podonos}. Comparison charts show ARCAS 290M in blue, commercial detectors in orange, and published baselines in grey.

\subsection{ASVspoof 5 Evaluation}\label{sec:asvspoof5}
ASVspoof~5 is the fifth edition of the automatic speaker verification spoofing challenge \cite{wang2024asvspoof5,wang2025asvspoof5db}. Its Track~1 evaluation set contains 680,774 records (138,688 bona fide, 542,086 spoofed) built from Multilingual LibriSpeech. It covers 16 unseen attacks, including TTS, voice conversion, and the adversarial filters Malafide and Malacopula, and every attack appears under 12 conditions: uncoded, Opus, AMR, Speex, EnCodec, MP3, M4A, MP3 followed by EnCodec, three narrowband codecs, and real telephone and Bluetooth devices. All 680,774 records were scored.

\begin{figure}[!htb]\centering
\includegraphics[width=\columnwidth]{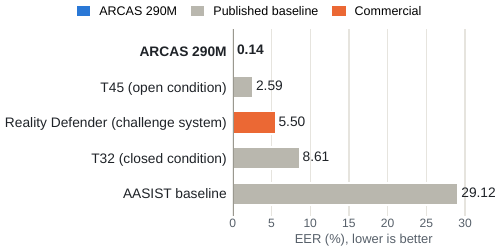}
\caption{Neural Defend ARCAS 290M and selected ASVspoof~5 Track~1 evaluation results (EER, \%). T45 and T32 are the top-ranked open- and closed-condition submissions; AASIST is the official baseline; the Reality Defender value is from its challenge paper \cite{zhu2024learnfromreal}. Values are contextual, not a protocol-matched ranking. ARCAS official Track~1 metrics: minDCF 0.00361, actDCF 0.00644, $C_\mathrm{llr}$ 0.01486 bits. Source: ASVspoof~5 workshop paper, Table~4 \cite{wang2024asvspoof5}.}\label{fig:sota_asvspoof5}
\end{figure}

Figure~\ref{fig:sota_asvspoof5} compares the 0.14\% EER of ARCAS 290M with the best open- and closed-condition challenge submissions and the AASIST baseline. With the official evaluation package, ARCAS reaches minDCF 0.00361, against 0.0750 for the best open-condition submission. The endpoint returns an uncalibrated score, mapped to $\log((1-s_i)/s_i)$ for the package, so actDCF (0.00644) and $C_\mathrm{llr}$ (0.01486 bits) describe the system as deployed. Challenge systems were trained under the closed rules (ASVspoof~5 training data only) or the open rules (external data allowed), so the figure is contextual. ARCAS 290M is below the lowest reported challenge EER, by a factor of more than 18.

\begin{figure}[!htb]\centering
\includegraphics[width=\columnwidth]{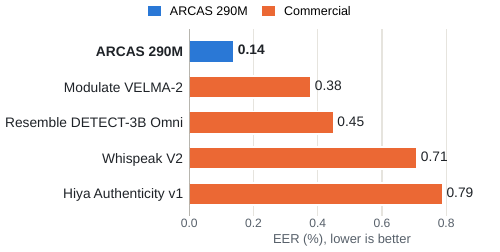}
\caption{Neural Defend ARCAS 290M and commercial detectors on ASVspoof~5 evaluation data (EER, \%). Speech DF Arena values are from its ASV2024-Eval column, accessed November 2026 \cite{dowerah2026dfarena}; the arena's evaluation subset is not documented, so these values are indicative. Reality Defender: Track~1 evaluation EER from its challenge paper \cite{zhu2024learnfromreal}.}\label{fig:commercial_asvspoof5}
\end{figure}

Commercial detectors have also been evaluated on ASVspoof~5. The Speech DF Arena leaderboard, run by its academic maintainers, reports EERs on the ASVspoof~5 evaluation data of 0.38\% for Modulate VELMA-2, 0.45\% for Resemble DETECT-3B Omni, 0.71\% for Whispeak V2, and 0.79\% for Hiya \cite{dowerah2026dfarena}, and Reality Defender's challenge system reached 5.5\% EER on the Track~1 evaluation set \cite{zhu2024learnfromreal}. ARCAS 290M, at 0.14\%, has the lowest EER of these systems (Figures~\ref{fig:commercial_asvspoof5} and~\ref{fig:sota_asvspoof5}). The arena does not state whether it uses all 680,774 Track~1 evaluation records, so this comparison is indicative.

ARCAS 290M makes 521 errors across 680,774 records: 302 missed fakes and 219 false alarms. Recall is 99.94\% and specificity 99.84\%. Accuracy stays between 99.69\% and 99.99\% across all 12 conditions (Table~\ref{tab:asvspoof5_conditions}); the weakest is MP3 followed by EnCodec (99.69\%, EER 0.35\%), which approximates audio re-shared across platforms. Fifteen of 16 attacks are detected at 99.9\% or higher, including both adversarial filters.
\input{tables/asvspoof5_conditions_inline}

The residual errors are concentrated rather than random. The pre-trained YourTTS attack accounts for 146 of the 302 misses (99.58\% recall on 34,414 records), and by condition 111 misses follow MP3 plus EnCodec, 72 narrowband Opus, and 58 EnCodec. False alarms are spread across conditions without a dominant cause. The result should be read as robust to transmission, with double compression and one TTS attack as the measurable weak points. In practice, a voice clone delivered over WhatsApp, a mobile call, or a Bluetooth headset is caught as reliably as a clean file, with about one error in every 1,300 recordings.

\subsection{SpeechFake Evaluation}\label{sec:speechfake}
SpeechFake, published at ACL 2025, is a large-scale multilingual speech-deepfake dataset built with about 40 generation tools \cite{speechfake2025}. Its release defines test sets as experiment lists rather than as a single fixed evaluation split; we score the union of all official test lists, 906,068 records (117,841 real, 788,227 fake) from 31 open-source generators in 13 languages. The ten commercial-API subsets are unreleased and not evaluated, and no official test list contains Hindi. Released counts differ slightly from the paper; we use the released files.

\begin{figure}[!htb]\centering
\includegraphics[width=\columnwidth]{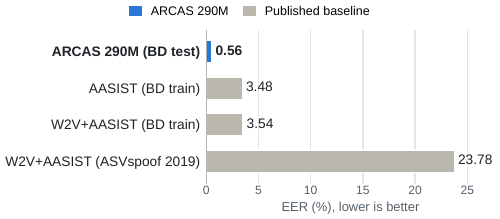}
\caption{Neural Defend ARCAS 290M and selected published SpeechFake results (EER, \%). All values are on the bilingual (BD) test list. Source: SpeechFake, Table~2 \cite{speechfake2025}.}\label{fig:sota_speechfake}
\end{figure}

Figure~\ref{fig:sota_speechfake} compares ARCAS 290M with AASIST and W2V+AASIST trained on SpeechFake's bilingual training split and with W2V+AASIST trained on ASVspoof 2019, all on the bilingual (BD) test list. On that list alone (346,307 records: 309,065 fake, 37,242 real), ARCAS reaches 0.556\% EER (95\% CI 0.523--0.588), 99.70\% accuracy, and AUROC 0.9998; English and Chinese sub-lists give 0.576\% and 0.485\%. Pooled over all lists, EER is 0.23\%.

ARCAS 290M reaches 99.92\% recall and 99.58\% specificity. TTS is detected at 100.00\% (2 of 434,878 missed) and voice conversion at 100.00\% (10 of 273,076 missed). EER is 0.44\% for English and 0.35\% for Chinese, and the other 11 languages are at 0.00--0.10\%, except Icelandic, which has too few real records.

Misses concentrate in one generator family. Of the 665 misses, 635 are BigVGAN vocoder re-syntheses (94.96\% recall on 12,601 records) and 18 are StyleMelGAN. Vocoder re-synthesis keeps the speaker and words and leaves few artifacts, making it the one measurable gap. False alarms (0.42\% of real speech) are mostly English (275) and Chinese (205). For a platform handling voice notes in these 13 languages, about one fake in 1,200 would pass unflagged.

\subsection{ShiftySpeech Evaluation}\label{sec:shiftyspeech}
ShiftySpeech tests whether detectors hold up under distribution shift \cite{shiftyspeech2025}. Real speech from audiobooks, podcasts, YouTube, AISHELL Mandarin read speech, and JSUT Japanese read speech is re-synthesized by 12 neural vocoders (APNet2, BigVGAN, BigVSAN, HiFi-GAN, iSTFTNet, MB-MelGAN, Parallel WaveGAN, Style-MelGAN, UnivNet in two sizes, Vocos, and WaveGrad). The same vocoders are applied to Common Voice, LJSpeech, VoxCeleb, LibriTTS train-clean-360, and VCTK, and the release adds TTS fakes from XTTS, VITS, FastPitch, Glow-TTS, Grad-TTS, and YourTTS. We score all 106 archives of the released test split: 2,781,458 records (58,806 real, 2,722,652 fake, of which 2,466,712 are vocoder re-syntheses and 255,940 TTS). Only the five domains listed first contain real speech.

\begin{figure}[!htb]\centering
\includegraphics[width=\columnwidth]{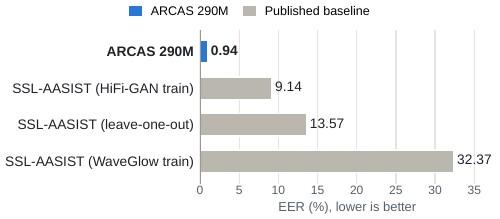}
\caption{Neural Defend ARCAS 290M pooled EER and author-reported SSL-AASIST average EER on ShiftySpeech (\%). SSL-AASIST values average over test vocoders and domains and depend on the training vocoder. Source: ShiftySpeech, Tables~5--6 \cite{shiftyspeech2025}.}\label{fig:sota_shiftyspeech}
\end{figure}

Figure~\ref{fig:sota_shiftyspeech} compares the 0.94\% pooled EER of ARCAS 290M with the authors' SSL-AASIST detector trained on WaveFake vocoders. The authors report average EER over test vocoders and domains, from 9.14\% (trained on HiFi-GAN) to 32.37\% (trained on WaveGlow), and 13.57--14.79\% in a leave-one-vocoder-out setting. The metrics differ, so the comparison is contextual.

ARCAS 290M reaches 99.33\% recall and 98.83\% specificity (AUROC 0.9995). All six TTS systems are detected at 100.00\% (11 of 255,940 missed). Among vocoders, recall ranges from 95.05\% for APNet2 and 97.57\% for BigVGAN to 99.56\% for UnivNet, MB-MelGAN, Style-MelGAN, and Parallel WaveGAN. By source domain, EER is 0.02\% for JSUT, 0.28\% for podcasts, 0.38\% for audiobooks, and 0.59\% for AISHELL, but 5.42\% for YouTube, where recall falls to 93.64\% and specificity to 95.88\%. By language, EER is 1.05\% for English, 0.25\% for Chinese, and 0.02\% for Japanese.

The shift that matters is the source audio, not the language or the vocoder. Of the 18,197 misses, 11,975 come from YouTube-sourced audio, and 646 of the 690 false alarms are real YouTube clips. By generator, misses are led by BigVGAN (4,883), APNet2 (3,149), and Vocos (2,441). Already-compressed, noisy web audio is the weak condition, consistent with the in-the-wild results below. For content platforms, clips re-uploaded from video sites deserve a second check, while podcast, audiobook, and read speech is screened almost perfectly.

\subsection{CodecFake+ Evaluation}\label{sec:codecfake}
CodecFake+ studies speech produced by neural-codec language models \cite{codecfakeplus2025}. It has two evaluation sets. CoSG contains speech from 17 codec-LM TTS systems, including VALL-E, MaskGCT, and NaturalSpeech~2 and~3; its released label file lists 1,797 records (866 real, 931 fake), 16 more real records than the paper. The CoRS evaluation set re-synthesizes real VCTK speech of speakers p227 and p228 through 31 neural codecs: 24,160 records (755 real, 23,405 fake). Four codec versions absent from the official label file were not scored.

\begin{figure}[!htb]\centering
\includegraphics[width=\columnwidth]{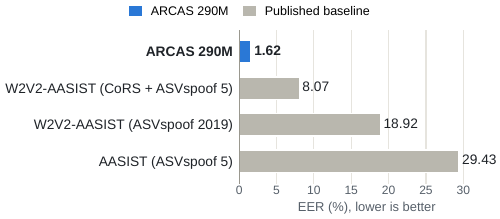}
\caption{Neural Defend ARCAS 290M and selected published CodecFake+ CoSG results (EER, \%). The paper reports CoSG on 850 real records, against 866 in the released labels. Values are contextual. Source: CodecFake+, Table~VI \cite{codecfakeplus2025}.}\label{fig:sota_codecfake_cosg}
\end{figure}

\begin{figure}[!htb]\centering
\includegraphics[width=\columnwidth]{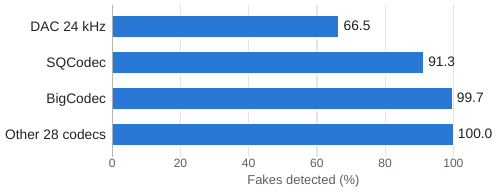}
\caption{Neural Defend ARCAS 290M recall on CodecFake+ CoRS evaluation fakes for the three weakest codecs and the remaining 28 codecs (755 fakes per codec) \cite{codecfakeplus2025}.}\label{fig:sota_codecfake_cors}
\end{figure}

Figure~\ref{fig:sota_codecfake_cosg} compares the 1.62\% CoSG EER of ARCAS 290M with W2V2-AASIST and AASIST trained on ASVspoof data, with or without CoRS. The strongest baseline, trained with CoRS and ASVspoof~5, reaches 8.07\%. CoRS EERs are reported only graphically and are not compared; Figure~\ref{fig:sota_codecfake_cors} instead shows ARCAS recall by codec.

The two sets fail differently. On CoSG, 16 of 17 systems are detected at 100\% and GPST at 93.3\% (2 of 30 missed), but 16 real records (1.8\%) are flagged as fake, nine of them short UniAudio voice prompts from demo pages. On CoRS, 28 of 31 codecs are detected at 100\%, while DAC at 24~kHz is detected at 66.5\% (253 of 755 missed) and 4.9\% of real VCTK records are flagged as fake.

Codec-LM TTS output is detected almost completely; residual risks are DAC re-synthesis (a third passes as real) and false alarms on short or re-synthesized real speech. Voices from the newest codec-based TTS systems, such as VALL-E and MaskGCT, are therefore already covered.

\subsection{ML-ITW Evaluation}\label{sec:mlitw}
ML-ITW is an evaluation-only collection of real and fake speech from seven social platforms in 14 languages, covering 180 public figures \cite{mlitw2026}. We score its full evaluation split: 24,529 records (18,168 real, 6,361 fake). The generators are unknown and largely commercial, and every record has passed through platform compression, often with music, noise, or editing.

\begin{figure}[!htb]\centering
\includegraphics[width=\columnwidth]{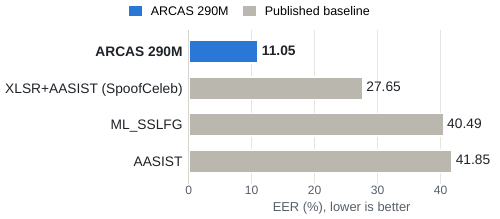}
\caption{Neural Defend ARCAS 290M and selected published ML-ITW results (EER, \%), all zero-shot on the same evaluation split. Values are contextual, not a protocol-matched ranking. Source: ML-ITW, Tables~3--4 \cite{mlitw2026}.}\label{fig:sota_mlitw}
\end{figure}

Figure~\ref{fig:sota_mlitw} compares the 11.05\% EER of ARCAS 290M with the best detectors reported by the authors. All systems are evaluated on the same split and none was trained on ML-ITW. The strongest plotted baseline, XLSR+AASIST trained on SpoofCeleb, reaches 27.65\%; off-the-shelf detectors score 40.49\% or higher.

ARCAS 290M reaches 78.84\% recall and 95.74\% specificity (AUROC 0.938). By language and platform (Table~\ref{tab:mlitw_subsets}), recall ranges from 97.0\% for Spanish (EER 2.4\%) to 22.7\% for Russian (44 fakes; EER 28.5\%); Portuguese (56.5\%) and Turkish (60.2\%) are also weak, while Hindi reaches 79.1\% (EER 15.2\%). By platform, recall ranges from 92.8\% on TikTok to 66.3\% on RedNote.

The ML-ITW misses are among the most confident in the study: the median fake score of a missed fake is 0.007, and 72\% score below 0.1. False alarms cluster on Bilibili (546 of 774; 12.8\% of real Bilibili records). The result is substantially above published zero-shot detectors but identifies language and platform as the main residual factors. For social-media monitoring, ARCAS 290M catches about four in five circulated deepfakes without tuning, more than any published detector, and its false alarms concentrate on one platform that can be checked during onboarding.
\input{tables/mlitw_subsets_inline}

\subsection{EchoFake Evaluation}\label{sec:echofake}
EchoFake contains AI-generated fakes from 11 recent zero-shot voice-cloning systems released 2023--2025, including F5-TTS, LLaSA, MaskGCT, and StyleTTS2 \cite{echofake2025}. Real and fake speech are also replayed through loudspeakers and re-recorded on phones, laptops, and earbuds in varied rooms and at varied distances. We score the closed-set (5,991) and open-set (25,600) evaluation splits, 31,591 records: 7,900 real, 7,900 fake, 7,898 replayed real, and 7,893 replayed fake. Replayed records keep their original label; the public open set has 400 fewer records than the paper's 32,000.

\begin{figure}[!htb]\centering
\includegraphics[width=\columnwidth]{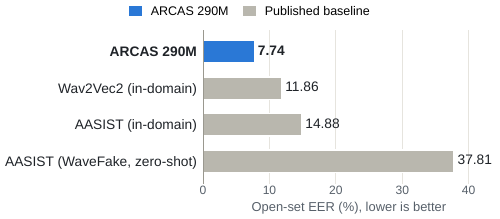}
\caption{Neural Defend ARCAS 290M and selected published EchoFake open-set results (EER, \%). In-domain baselines are trained on EchoFake; the WaveFake-trained AASIST is zero-shot. Source: EchoFake, Tables~3--4 \cite{echofake2025}.}\label{fig:sota_echofake}
\end{figure}

Figure~\ref{fig:sota_echofake} compares the 7.74\% open-set EER of ARCAS 290M with in-domain and zero-shot baselines. The open set adds unseen speakers, TTS systems, and replay conditions. On the closed set, the in-domain wav2vec~2.0 baseline is stronger (0.27\% against 10.83\%).

ARCAS 290M reaches 78.74\% recall with 99.65\% specificity. On original audio, 99.84\% of fakes are detected and 99.94\% of real speech passes (EER 0.15\%). After replay, 57.62\% of fakes are detected, while 99.37\% of replayed real speech passes (EER 16.29\%).

Replay accounts for 3,345 of the 3,358 misses. Detection of replayed fakes is lowest on wired earbuds (38.1\%), a Samsung Galaxy A54 (43.8\%), and at 50~cm distance (44.0\%). The most natural generators lose the most (original and replayed pooled): LLaSA is detected at 60.7\%, MaskGCT at 63.8\%, and StyleTTS2 at 66.5\%, while the older SpeechT5 stays at 96.2\%. Protection is strongest on digitally captured audio, such as call recordings and uploads, rather than loudspeaker re-recordings.

\subsection{HQ-MPSD Evaluation}\label{sec:hqmpsd}
HQ-MPSD is a multilingual partial-deepfake benchmark built from Multilingual LibriSpeech in eight languages \cite{hqmpsd2025}. Real utterances are partly re-spoken with XTTSv2 and spliced at word boundaries with loudness matching and cross-fades; some partial fakes then get background noise, echo, or both. No English test split is published, so we score all clips in all eight languages: 204,038 records (51,715 real; 51,715 fully fake; 51,715 clean partial; 48,893 partial with noise or echo). The English archive lacks frame labels and part of its noisy partial set, so English fake-share and noise slices are not reported.

\begin{figure}[!htb]\centering
\includegraphics[width=\columnwidth]{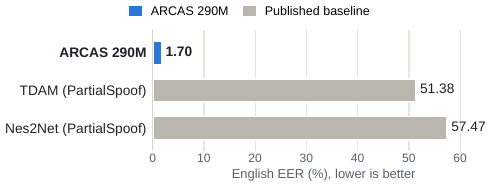}
\caption{Neural Defend ARCAS 290M and selected published HQ-MPSD English results (EER, \%). Both baselines are trained on PartialSpoof and evaluated zero-shot. Values are contextual. Source: HQ-MPSD, Sec.~IV-B \cite{hqmpsd2025}.}\label{fig:sota_hqmpsd}
\end{figure}

Figure~\ref{fig:sota_hqmpsd} compares the 1.7\% English EER of ARCAS 290M with TDAM and Nes2Net trained on PartialSpoof and evaluated zero-shot. TDAM trained on HQ-MPSD's own English split reaches 0.29\% in-domain, but 27.49--36.13\% on the seven other languages, where ARCAS reaches 2.4--14.8\% (Table~\ref{tab:hqmpsd_language}).
\input{tables/hqmpsd_language}

ARCAS 290M reaches 83.51\% recall and 94.49\% specificity. Fully fake records are detected at 100\%, clean partial fakes at 83.2\%, and partial fakes with background noise, room echo, or both at 78.3\%, 61.1\%, and 56.3\%. Partial fakes covering under 25\% of the clip are detected at 19.1\% (758 records).

Two separate problems appear. Recording degradation compounds partial-fake difficulty: 16,400 of the 25,113 misses carry room echo or noise. Independently, real Polish speech is flagged as fake 22.3\% of the time (2,203 of 2,849 false alarms), while every other language passes 95.6\% or more; Polish is the only language-linked false-alarm problem within HQ-MPSD. Multilingual deployments can expect most partial edits in clean audio to be caught, and Polish-language deployments should validate false alarms on local speech.

\subsection{LlamaPartialSpoof Evaluation}\label{sec:llamapartialspoof}
LlamaPartialSpoof is released for evaluation only and tests detectors on short, natural-sounding edits that keep the speaker's identity \cite{llamapartialspoof2024}. An LLM rewrites part of a real LibriTTS transcript, and the new words are spoken in the original speaker's voice by one of six systems (CosyVoice, ElevenLabs, GPT-SoVITS, LJ-JETS, XTTS-v2, YourTTS) and joined by cut-and-paste, cross-fade, or overlap-add. We score the full v1.0.b release (parts a and b): 140,616 records, comprising 10,573 real, 33,461 fully fake, and 96,582 partial (32,194 per joining method).

\begin{figure}[!htb]\centering
\includegraphics[width=\columnwidth]{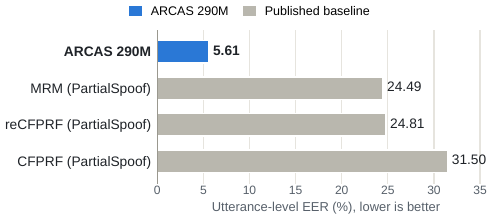}
\caption{Neural Defend ARCAS 290M and selected published LlamaPartialSpoof results (utterance-level EER, \%), all zero-shot. Values are contextual. Sources: LlamaPartialSpoof, Table~III \cite{llamapartialspoof2024}; Robust Localization, Table~III \cite{robustloc2025}.}\label{fig:sota_llamapartialspoof}
\end{figure}

Figure~\ref{fig:sota_llamapartialspoof} compares the 5.61\% utterance-level EER of ARCAS 290M with detectors trained on PartialSpoof and evaluated zero-shot, the strongest of which reaches 24.49\%, more than four times the ARCAS error. Our population includes part~b of release v1.0.b; the authors' own tables also cover all three joining methods, so the populations are closely aligned \cite{llamapartialspoof2024}.

ARCAS 290M reaches 81.46\% recall with no false alarms. Detection falls with the share of the clip that is fake (Table~\ref{tab:partial_share}): 99.97\% for whole-clip fakes, 95.08\% at 50\% or more, 65.17\% at 25--50\%, 26.73\% at 10--25\%, and 4.00\% below 10\%. By joining method, recall is 77.3\% (cut-and-paste), 75.9\% (cross-fade), and 71.9\% (overlap-add); by system, it ranges from 66.3\% (LJ-JETS) to 88.6\% (YourTTS), with ElevenLabs at 79.7\%.
\input{tables/partial_share_inline}

Detection depends on how much of the file is fake, not on which system produced it: 24,098 of the 24,109 misses are partial fakes. The misses are confident (median fake score 0.04; 59\% below 0.1), consistent with a single file-level verdict in which the real speech outweighs a short fake phrase. For compliance and evidence review, ARCAS 290M reliably flags clips that are mostly synthetic, with no false alarms on 10,573 real recordings, while clips with a single rewritten phrase should go to segment-level review.

\subsection{IndicSynth Evaluation}\label{sec:indicsynth}
IndicSynth, published at ACL 2025, contains synthetic speech in 12 Indian languages from FreeVC24 (voice conversion), XTTS-v2 (multilingual voice-cloning TTS), and VITS (TTS, Bengali only) \cite{indicsynth2025}. The release has no test split and no real speech. We score all Hindi (205,938), Tamil (282,312), Bengali (83,448), and Telugu (169,896) records (741,594 fakes in total) and report recall only.

\begin{figure}[!htb]\centering
\includegraphics[width=\columnwidth]{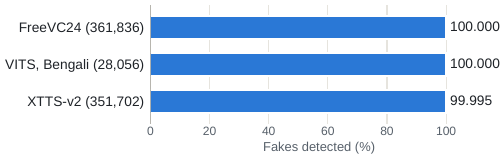}
\caption{Neural Defend ARCAS 290M recall on IndicSynth fakes by generation tool (\%; record counts in parentheses), Hindi, Tamil, Bengali, and Telugu \cite{indicsynth2025}.}\label{fig:sota_indicsynth}
\end{figure}

Figure~\ref{fig:sota_indicsynth} reports recall by generation tool. The authors report 8--93\% EER for ASVspoof 2019-trained AASIST, AASIST-L, and RawNet2 on 16,000-record subsets paired with IndicSUPERB real speech \cite{indicsynth2025}; those numbers use a real class and are not comparable to fake-only recall.

ARCAS 290M detects 741,577 of 741,594 fakes (99.998\%; 95\% Wilson interval 99.996--99.999). FreeVC24 (361,836) and VITS (28,056) are detected completely, and XTTS-v2 at 99.995\% (17 of 351,702 missed). By language, misses are 3 in Tamil, 4 in Hindi, 4 in Telugu, and 6 in Bengali; by speaker sex, 10 female and 7 male.

Every miss is an XTTS-v2 clip, and most are borderline: 9 of the 17 have fake scores between 0.36 and 0.50, and 7 score below 0.1. The four evaluated languages show no language- or tool-specific weakness. With no real speech, this result does not measure false alarms on Indian-language audio. For Indian-language markets, voice clones in Hindi, Tamil, Bengali, and Telugu are detected almost without exception: 17 misses in 741,594 clips.

\subsection{PartialEdit Evaluation}\label{sec:partialedit}
PartialEdit changes one word in real VCTK sentences with a neural speech editor, mimicking the subtle edits used to alter recorded statements or interviews \cite{partialedit2025}. Two editors are released, VoiceCraft and SSR-Speech; only the edited word is generated and pasted back into the original recording, so the file keeps the speaker, prosody, and recording conditions of the original. Each edited set also has a codec variant in which the whole file has passed through the editor's neural codec. We score the official 70-speaker evaluation split of release v1.1: 107,454 fakes. With no real speech, only recall is reported.

\begin{figure}[!htb]\centering
\includegraphics[width=\columnwidth]{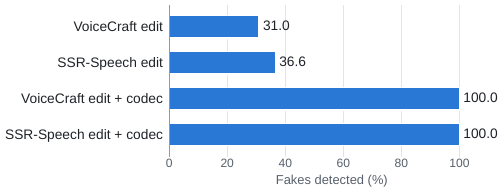}
\caption{Neural Defend ARCAS 290M recall on PartialEdit v1.1 evaluation fakes by subset (26,841 VoiceCraft and 26,886 SSR-Speech records per variant) \cite{partialedit2025}.}\label{fig:sota_partialedit}
\end{figure}

Figure~\ref{fig:sota_partialedit} reports recall by subset. Author-reported baselines use EER with real VCTK speech and are not comparable to fake-only recall; for context, XLSR-SLS trained on PartialSpoof reaches 23.72\% EER on PartialEdit and 12.95\% on PartialEdit-Codec \cite{partialedit2025}.

ARCAS 290M reaches 66.91\% recall (95\% Wilson interval 66.62--67.19). Single-word edits without the codec pass are detected at 31.0\% (VoiceCraft) and 36.6\% (SSR-Speech), but the same edits are detected at 100\% when the whole file has passed through the codec. By edited share of the clip, recall is 2.4\% below 10\%, 24.6\% at 10--25\%, 62.6\% at 25--50\%, and 92.8\% at 50\% or more (471 records).

The plain-versus-codec contrast indicates that the detector responds to file-wide codec artifacts, not to the edit itself. Misses are split evenly between editors (VoiceCraft 18,509; SSR-Speech 17,052) and are confident (median fake score 0.004; 77\% below 0.1). Where a single altered word matters, as in recorded statements, ARCAS 290M should be paired with segment-level review.

\subsection{Cross-benchmark analysis}\label{sec:cross_benchmark}
Table~\ref{tab:main_metrics} follows the same decreasing pooled-accuracy sequence as the result subsections. It consolidates accuracy, EER, AUROC, $F_1$, precision, recall, specificity, APCER, and BPCER for all evaluated populations. It provides a shared operating view, but is not a single leaderboard: real-to-fake ratios differ widely, and PartialEdit and IndicSynth contain no real speech.

Table~\ref{tab:sota_comparison} sets each result beside the strongest baseline published for that benchmark; ARCAS 290M has the lower error on every benchmark with a zero-shot comparison.
\input{tables/sota_comparison}
\input{tables/main_metrics}
\begin{figure*}[!t]\centering
\includegraphics[width=\textwidth]{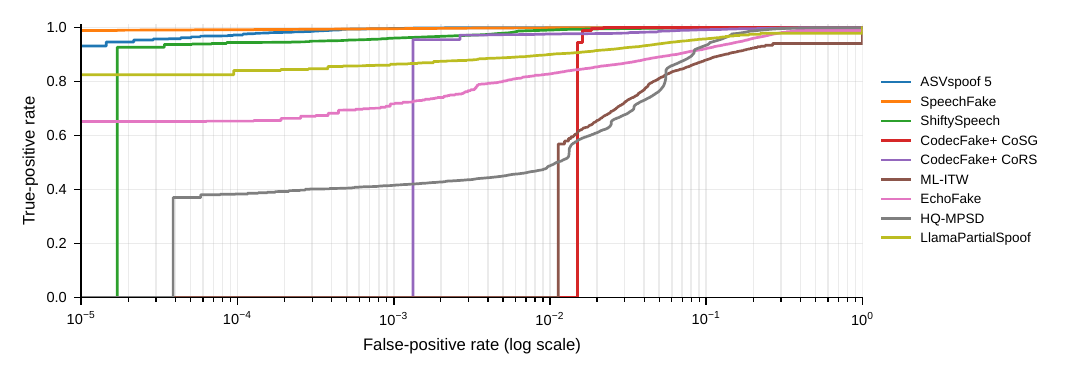}
\caption{Neural Defend ARCAS 290M low-false-positive-rate ROC curves across evaluated benchmark populations. Benchmark sources are cited in the corresponding subsections.}\label{fig:low_fpr_roc}
\end{figure*}
\begin{figure*}[!t]\centering
\includegraphics[width=\textwidth]{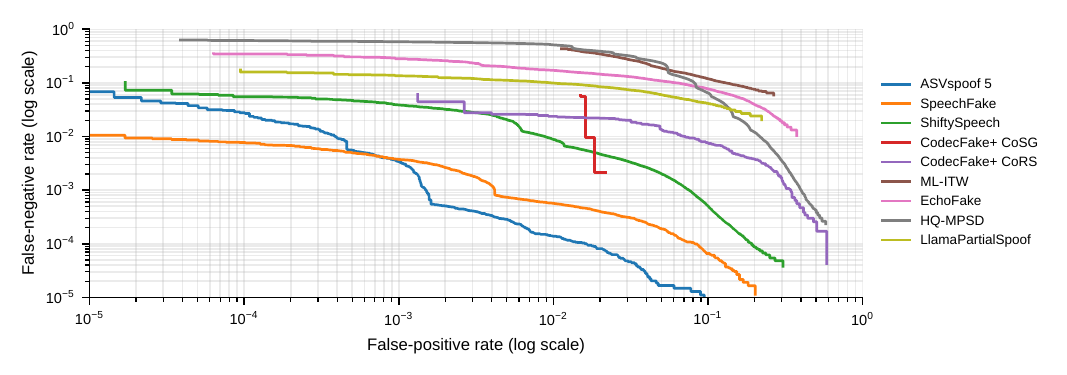}
\caption{Neural Defend ARCAS 290M detection-error tradeoff (DET) curves across the evaluated benchmark populations cited in the corresponding Results subsections.}\label{fig:det_curves}
\end{figure*}

Accuracy falls from 99.92\% to 82.85\% across benchmarks with real speech, and the fall follows the type of fake rather than the generator. Across the five benchmarks of whole-file synthetic or re-synthesized speech (ASVspoof~5, SpeechFake, ShiftySpeech, CoSG, and IndicSynth), 4,776,307 of 4,795,490 fakes are detected (99.60\%). Partial, replayed, and in-the-wild audio occupy the lower rows.

Coverage and stability held across the whole evaluation. ARCAS 290M returned a verdict for every one of the 5,644,079 records, including MP3, FLAC, WAV, and raw-byte audio at their original sample rates, and no file was declined, so every reported rate applies to the complete test set. One frozen model and one default threshold produced all results, from studio read speech to telephone codecs, loudspeaker replay, and social-media uploads, and canary recordings re-scored throughout the run returned identical outputs. A customer therefore receives the same behavior reported here without per-benchmark tuning.

Accuracy and EER can rank benchmarks differently. LlamaPartialSpoof has the lowest accuracy but a lower EER than HQ-MPSD, EchoFake, or ML-ITW: with a real-to-fake ratio of about 1:12 and no false alarms, its accuracy is driven almost entirely by missed partial fakes. Residual-error type also changes by benchmark: missed fakes dominate EchoFake, ML-ITW, HQ-MPSD, and LlamaPartialSpoof, while CoSG and CoRS show comparatively larger BPCER. False alarms are low on studio speech (BPCER 0.00--0.42\%) and rise on re-synthesized, platform-processed, or Polish real speech (1.17--5.51\%). For buyers, the practical reading of Table~\ref{tab:main_metrics} is to match each row to the audio a deployment will see: contact centers and voice-note platforms resemble the upper rows, while review of edited recordings or audio re-captured from a loudspeaker resembles the lower rows.

In the three weakest regions, missed fakes are confident, with median fake scores of 0.004--0.04, so changing the operating point would not recover them. Figure~\ref{fig:low_fpr_roc} compares low-false-positive-rate ROC behavior across benchmarks with real speech, and Figure~\ref{fig:det_curves} shows the full detection-error tradeoff: ASVspoof~5, SpeechFake, and ShiftySpeech stay close to the origin across the whole range, while ML-ITW, EchoFake, HQ-MPSD, and LlamaPartialSpoof keep a high miss rate until the false-alarm rate becomes large. The ROC view is most useful where false alarms carry a high review cost: curves nearer the upper-left retain fake-speech sensitivity while constraining real-speech errors. Curves step through each distinct four-decimal score; benchmarks with few real records (CoRS: 755) cannot resolve FPR below $1/n$.
\FloatBarrier

%% file: tables/asvspoof5_conditions_inline.tex
\begin{table}[H]
\centering
\caption{ASVspoof 5 Track 1 evaluation: accuracy and EER (\%) by transmission condition.}
\label{tab:asvspoof5_conditions}
\footnotesize
\setlength{\tabcolsep}{3.5pt}
\begin{tabular}{@{}lp{0.40\columnwidth}rr@{}}
\toprule
Condition & Typical real-world channel & Acc. & EER \\
\midrule
No compression & Original file & 99.99 & 0.05 \\
Opus, wideband & WhatsApp, Zoom, Discord & 99.96 & 0.14 \\
AMR, wideband & Mobile call (VoLTE) & 99.93 & 0.17 \\
Speex, wideband & Older VoIP applications & 99.94 & 0.04 \\
EnCodec & Neural compression & 99.81 & 0.28 \\
MP3 & Shared audio files & 99.99 & 0.03 \\
M4A / AAC & Phone memos, video platforms & 99.97 & 0.11 \\
MP3 then EnCodec & Audio re-shared across platforms & 99.69 & 0.35 \\
Opus, narrowband & Low-bandwidth call & 99.79 & 0.29 \\
AMR, narrowband & 2G/3G telephone call & 99.96 & 0.13 \\
Speex, narrowband & Low-quality VoIP & 99.92 & 0.10 \\
Real devices & Bluetooth or cable transmission & 99.97 & 0.10 \\
\bottomrule
\end{tabular}
\end{table}

%% file: tables/mlitw_subsets_inline.tex
\begin{table}[H]
\centering
\caption{ML-ITW: recall, specificity, and EER (\%) by language and platform. Ukrainian, Hebrew, and Hungarian have fewer than 30 fakes (17, 13, 3) and are omitted.}
\label{tab:mlitw_subsets}
\scriptsize
\setlength{\tabcolsep}{2.5pt}
\begin{tabular}{@{}lrrr@{\hspace{0.9em}}lrr@{}}
\toprule
Language & Rec. & Spec. & EER & Platform & Rec. & Spec. \\
\midrule
Spanish & 97.0 & 98.8 & 2.4 & TikTok & 92.8 & 99.5 \\
Japanese & 82.0 & 98.9 & 7.2 & Bilibili & 91.8 & 87.2 \\
English & 82.5 & 97.5 & 9.1 & X & 85.4 & 98.5 \\
Chinese & 78.8 & 90.9 & 12.3 & Facebook & 78.8 & 97.1 \\
French & 85.2 & 95.4 & 14.8 & Douyin & 77.5 & 99.5 \\
Hindi & 79.1 & 99.7 & 15.2 & YouTube & 74.2 & 98.9 \\
Korean & 69.1 & 99.8 & 15.3 & RedNote & 66.3 & 98.1 \\
German & 76.7 & 95.3 & 18.6 & & & \\
Turkish & 60.2 & 93.7 & 26.1 & & & \\
Russian & 22.7 & 100.0 & 28.5 & & & \\
Portuguese & 56.5 & 98.4 & 28.7 & & & \\
\bottomrule
\end{tabular}
\end{table}

%% file: tables/hqmpsd_language.tex
\begin{table}[H]
\centering
\caption{HQ-MPSD EER (\%) by language. TDAM is trained on the HQ-MPSD English training split, so English is in-domain and the other languages are cross-lingual. Source for TDAM: HQ-MPSD, Table~II \cite{hqmpsd2025}.}
\label{tab:hqmpsd_language}
\footnotesize
\setlength{\tabcolsep}{4pt}
\begin{tabular}{@{}lrrrr@{}}
\toprule
Language & Recall & Spec. & ARCAS EER & TDAM EER \\
\midrule
English & 93.0 & 99.6 & 1.7 & 0.29 \\
German & 79.3 & 99.7 & 2.4 & 32.65 \\
Dutch & 79.7 & 100.0 & 2.8 & 32.57 \\
Portuguese & 69.6 & 100.0 & 3.2 & 30.72 \\
Italian & 79.7 & 98.5 & 4.7 & 27.49 \\
French & 81.4 & 98.6 & 5.2 & 29.47 \\
Spanish & 83.3 & 95.6 & 7.1 & 28.88 \\
Polish & 91.8 & 77.7 & 14.8 & 36.13 \\
\bottomrule
\end{tabular}
\end{table}

%% file: tables/partial_share_inline.tex
\begin{table}[H]
\centering
\caption{Recall (\%) on partial-fake datasets by share of the clip that is fake.}
\label{tab:partial_share}
\small
\begin{tabular}{lrr}
\toprule
Fake share of clip & LlamaPartialSpoof & PartialEdit \\
\midrule
Whole clip & 99.97 & \NA \\
50\% or more & 95.08 & 92.8 \\
25--50\% & 65.17 & 62.6 \\
10--25\% & 26.73 & 24.6 \\
Under 10\% & 4.00 & 2.4 \\
\bottomrule
\end{tabular}
\end{table}

%% file: tables/sota_comparison.tex
\begin{table}[!t]
\centering
\caption{Neural Defend ARCAS 290M versus the strongest author-reported baseline per benchmark (EER, \%). Values keep each source's metric and protocol, so rows are not a unified leaderboard. Setting: In-dom.\ = baseline trained on the benchmark's own data; Zero-shot = trained elsewhere. ShiftySpeech compares pooled EER with the authors' average over test vocoders and domains; CoSG uses 866 real records against the paper's 850; HQ-MPSD (EN) is English only; PartialEdit is fake-only for ARCAS (recall 66.91\%).}
\label{tab:sota_comparison}
\footnotesize
\setlength{\tabcolsep}{2.5pt}
\begin{tabular}{@{}>{\raggedright\arraybackslash}p{0.255\columnwidth}rr>{\raggedright\arraybackslash}p{0.34\columnwidth}l@{}}
\toprule
Benchmark & ARCAS & Prior & Prior method & Setting \\
\midrule
ASVspoof 5 & 0.14 & 2.59 & Team T45 \citep{wang2024asvspoof5} & Challenge \\
SpeechFake (BD) & 0.56 & 3.48 & AASIST \citep{speechfake2025} & In-dom. \\
ShiftySpeech & 0.94 & 9.14 & SSL-AASIST \citep{shiftyspeech2025} & Zero-shot \\
CodecFake+ CoSG & 1.62 & 8.07 & W2V2-AASIST \citep{codecfakeplus2025} & Zero-shot \\
ML-ITW & 11.05 & 27.65 & XLSR+AASIST \citep{mlitw2026} & Zero-shot \\
EchoFake (open) & 7.74 & 11.86 & Wav2Vec2+MLP \citep{echofake2025} & In-dom. \\
EchoFake (closed) & 10.83 & 0.27 & Wav2Vec2+MLP \citep{echofake2025} & In-dom. \\
HQ-MPSD (EN) & 1.7 & 0.29 & TDAM \citep{hqmpsd2025} & In-dom. \\
LlamaPartialSpoof & 5.61 & 24.49 & MRM \citep{llamapartialspoof2024} & Zero-shot \\
PartialEdit & \NA & 23.72 & XLSR-SLS \citep{partialedit2025} & Zero-shot \\
\bottomrule
\end{tabular}
\end{table}

%% file: tables/main_metrics.tex
\begin{table*}[t]
\centering
\caption{Neural Defend ARCAS 290M per-benchmark performance metrics (\%, except AUROC). APCER is the fake-speech error rate ($1-\mathrm{Recall}$) and BPCER is the real-speech error rate ($1-\mathrm{Specificity}$). Accuracy intervals are 95\% Wilson intervals. SpeechFake pools all official test lists; on the BD list alone (346,307 records) accuracy is 99.70\%, EER 0.556\%, and AUROC 0.9998. PartialEdit and IndicSynth contain no real speech; only recall is defined.}
\label{tab:main_metrics}
\scriptsize
\setlength{\tabcolsep}{3pt}
\begin{tabular}{lrrrrrrrrrr}
\toprule
Dataset & Records & Accuracy (95\% CI) & EER & AUROC & $F_1$ & Precision & Recall & Specificity & APCER & BPCER \\
\midrule
ASVspoof 5 & 680,774 & 99.92 (99.92--99.93) & 0.14 & 1.0000 & 99.95 & 99.96 & 99.94 & 99.84 & 0.06 & 0.16 \\
SpeechFake & 906,068 & 99.87 (99.86--99.88) & 0.23 & 1.0000 & 99.93 & 99.94 & 99.92 & 99.58 & 0.08 & 0.42 \\
ShiftySpeech & 2,781,458 & 99.32 (99.31--99.33) & 0.94 & 0.9995 & 99.65 & 99.97 & 99.33 & 98.83 & 0.67 & 1.17 \\
CodecFake+ CoSG & 1,797 & 99.00 (98.42--99.37) & 1.62 & 0.9920 & 99.04 & 98.31 & 99.79 & 98.15 & 0.21 & 1.85 \\
CodecFake+ CoRS & 24,160 & 98.52 (98.36--98.66) & 2.19 & 0.9968 & 99.23 & 99.84 & 98.63 & 95.10 & 1.37 & 4.90 \\
ML-ITW & 24,529 & 91.36 (91.00--91.70) & 11.05 & 0.9381 & 82.55 & 86.63 & 78.84 & 95.74 & 21.16 & 4.26 \\
EchoFake & 31,591 & 89.20 (88.85--89.53) & 8.50 & 0.9751 & 87.93 & 99.56 & 78.74 & 99.65 & 21.26 & 0.35 \\
HQ-MPSD & 204,038 & 86.30 (86.15--86.44) & 8.38 & 0.9712 & 90.10 & 97.81 & 83.51 & 94.49 & 16.49 & 5.51 \\
LlamaPartialSpoof & 140,616 & 82.85 (82.66--83.05) & 5.61 & 0.9819 & 89.78 & 100.00 & 81.46 & 100.00 & 18.54 & 0.00 \\
\midrule
IndicSynth (fake only) & 741,594 & \NA & \NA & \NA & \NA & \NA & 99.998 & \NA & 0.002 & \NA \\
PartialEdit (fake only) & 107,454 & \NA & \NA & \NA & \NA & \NA & 66.91 & \NA & 33.09 & \NA \\
\bottomrule
\end{tabular}
\end{table*}

%% file: sections/07_discussion.tex
\section{Discussion}
\begin{figure*}[!b]
\centering
\begin{minipage}{0.24\textwidth}\centering
\includegraphics[width=\linewidth]{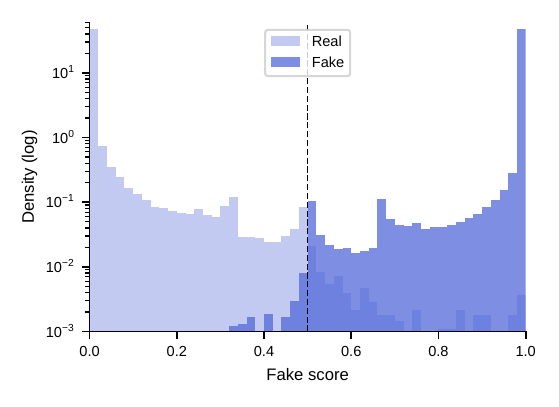}\\[-0.4em]\scriptsize ASVspoof 5
\end{minipage}\hfill
\begin{minipage}{0.24\textwidth}\centering
\includegraphics[width=\linewidth]{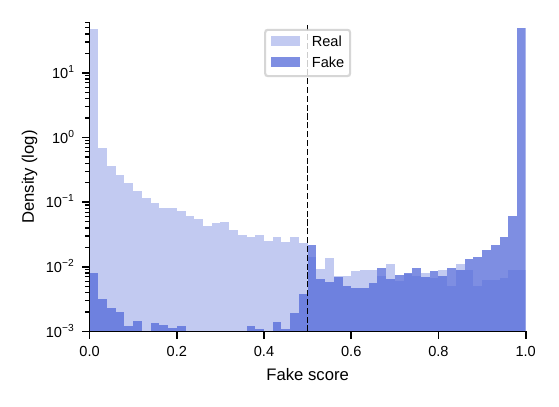}\\[-0.4em]\scriptsize SpeechFake
\end{minipage}\hfill
\begin{minipage}{0.24\textwidth}\centering
\includegraphics[width=\linewidth]{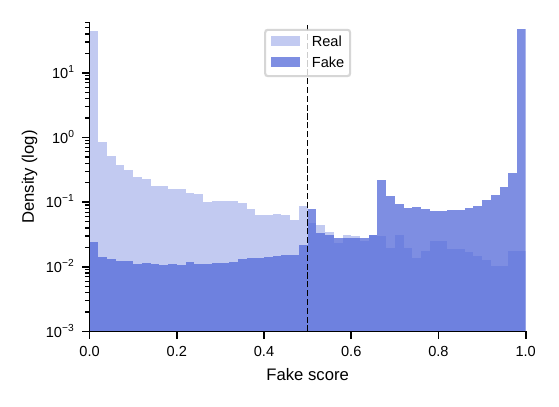}\\[-0.4em]\scriptsize ShiftySpeech
\end{minipage}\hfill
\begin{minipage}{0.24\textwidth}\centering
\includegraphics[width=\linewidth]{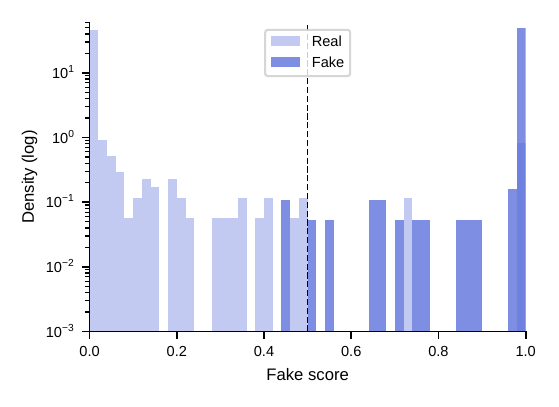}\\[-0.4em]\scriptsize CodecFake+ CoSG
\end{minipage}
\vspace{0.4em}
\begin{minipage}{0.24\textwidth}\centering
\includegraphics[width=\linewidth]{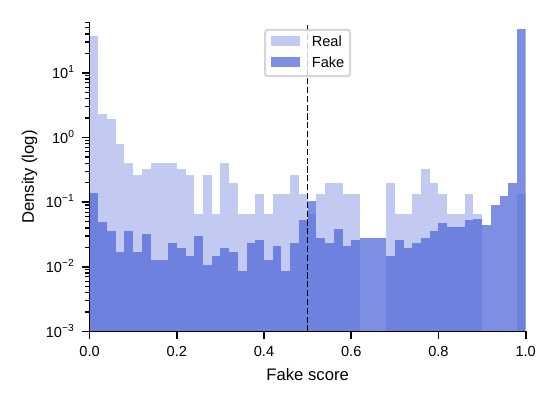}\\[-0.4em]\scriptsize CodecFake+ CoRS
\end{minipage}\hfill
\begin{minipage}{0.24\textwidth}\centering
\includegraphics[width=\linewidth]{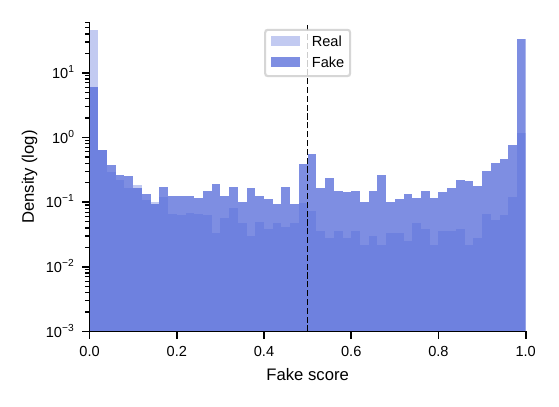}\\[-0.4em]\scriptsize ML-ITW
\end{minipage}\hfill
\begin{minipage}{0.24\textwidth}\centering
\includegraphics[width=\linewidth]{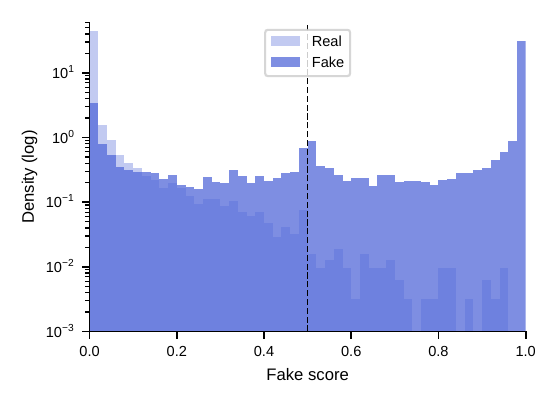}\\[-0.4em]\scriptsize EchoFake
\end{minipage}\hfill
\begin{minipage}{0.24\textwidth}\centering
\includegraphics[width=\linewidth]{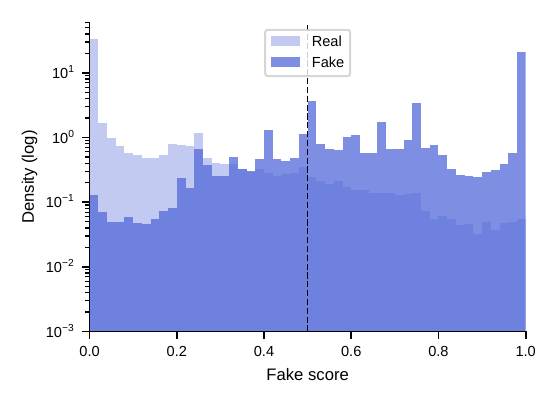}\\[-0.4em]\scriptsize HQ-MPSD
\end{minipage}
\vspace{0.4em}
\begin{minipage}{0.24\textwidth}\centering
\includegraphics[width=\linewidth]{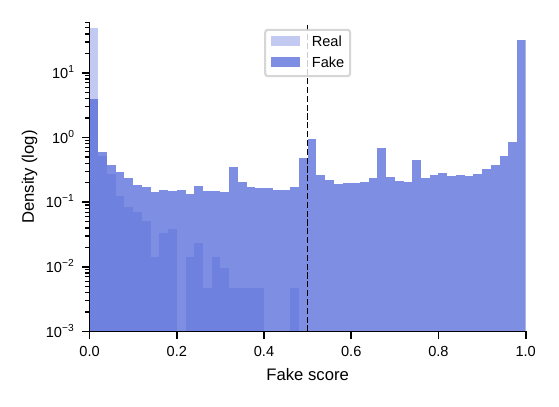}\\[-0.4em]\scriptsize LlamaPartialSpoof
\end{minipage}\hfill
\begin{minipage}{0.24\textwidth}\centering
\includegraphics[width=\linewidth]{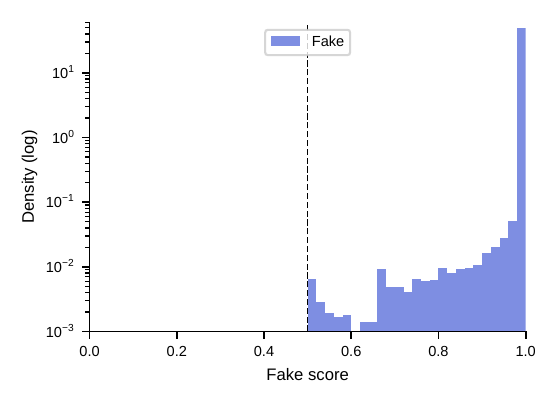}\\[-0.4em]\scriptsize IndicSynth (fake only)
\end{minipage}\hfill
\begin{minipage}{0.24\textwidth}\centering
\includegraphics[width=\linewidth]{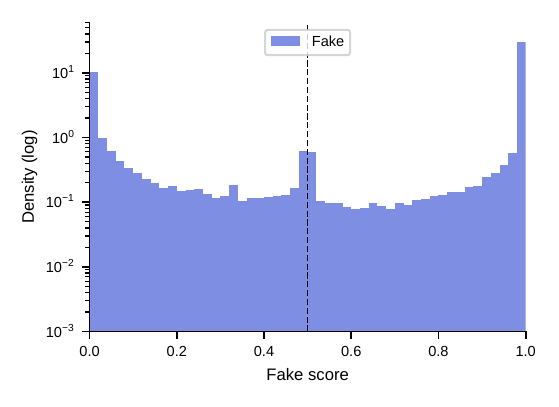}\\[-0.4em]\scriptsize PartialEdit (fake only)
\end{minipage}
\vspace{0.4em}
\caption{Neural Defend ARCAS 290M real-versus-fake score distributions by benchmark (log density; dashed line at 0.5).}
\label{fig:confidence_distributions}
\end{figure*}

\subsection{What the results mean in practice}
Results separate into two regimes. When the whole file is synthesized by TTS, voice-conversion, or codec-LM systems, ARCAS 290M separates real from fake almost perfectly: EER is 0.14--2.19\% on ASVspoof~5, SpeechFake, ShiftySpeech, and CodecFake+, and recall is 99.998\% on IndicSynth, holding across 12 transmission conditions and over 15 languages. When only part of the file is fake, the audio is replayed, or it comes from web platforms, detection depends on how much synthetic signal reaches the detector. There, fake share, acoustics, and devices matter more than the generator.

Figure~\ref{fig:confidence_distributions} shows the two regimes directly. On whole-file synthetic benchmarks the real and fake score distributions sit at opposite ends of the scale with little overlap, so the default threshold separates them cleanly. In the partial-fake, replay, and in-the-wild benchmarks, many fakes score near zero rather than near the 0.5 boundary, so the detector is confident, not uncertain, about these misses. The DET curves in Figure~\ref{fig:det_curves} show the same pattern across all nine benchmarks with real speech. For deployment, this means the default threshold needs no adjustment on whole-file synthetic audio. For partial, replayed, or web-sourced audio, lowering the threshold would raise false alarms faster than it recovers missed fakes, so these cases are better routed to segment-level analysis or human review, where an analyst can examine the suspect segment and the recording conditions. Routing rules can therefore be set by audio type rather than by generator, which is rarely known in practice.

Two observations point to the same mechanism. On PartialEdit, identical single-word edits are detected 31--37\% of the time, but 100\% after the whole file passes through a neural codec. On EchoFake, the same fakes are detected 99.84\% of the time before replay and 57.62\% after, while real speech passes more than 99\% of the time in both cases. The detector responds to signal-level artifacts spread across the file; when those artifacts are confined to a short span, or washed out by a loudspeaker and microphone, the file-level verdict follows the real portion. The same mechanism explains the weaker ML-ITW and ShiftySpeech YouTube results, where platform compression and background audio dilute the artifacts left by whole-file synthesis.

\subsection{Where to rely on ARCAS 290M and where to add review}
Under the stated conditions, the benchmark supports three operational statements. First, whole-file synthetic speech, including codec-compressed and telephone audio, is reliably detected; this covers voice-cloned calls and messages produced directly by TTS or voice-conversion tools. Second, partial manipulations are the main residual risk: a single replaced word or short phrase in genuine speech is usually not flagged, so use cases where such edits matter, such as evidence review or recorded statements, need segment-level analysis or human review. Third, over-the-air capture reduces detection: audio recorded through a microphone from a loudspeaker loses about 40 points of recall, especially with low-cost recording devices. Source and language effects add to these: already-compressed web audio (YouTube in ShiftySpeech, several platforms in ML-ITW), Russian, Portuguese, and Turkish social-media audio, and false alarms on Polish read speech are concrete targets for data collection and validation. IndicSynth shows high recall on four Indian languages, but without a real class it does not measure false alarms on Indian-language speech.

\subsection{How ARCAS 290M compares}
On every benchmark where the authors report a zero-shot baseline on an aligned population, ARCAS 290M has a lower EER, between 2.5 and more than 11 times below the strongest zero-shot baseline: 11.05\% against 27.65--50.03\% on ML-ITW, 7.74\% against 37.81--56.23\% on the EchoFake open set, 5.61\% against 24.49--57.18\% on LlamaPartialSpoof, and 1.62\% against 18.92--36.95\% on CodecFake+ CoSG. Against commercial detectors the picture is the same: ARCAS 290M leads the Podonos leaderboard and has the lowest ASVspoof~5 EER among the commercial systems on the Speech DF Arena. These comparisons are context rather than rank. The baselines are research models trained on one public corpus, whereas ARCAS is a production system; some populations differ slightly; and in-domain baselines trained on a benchmark's own data can outperform ARCAS on that benchmark, as on the EchoFake closed set (0.27\% against 10.83\%) and HQ-MPSD English (0.29\% against 1.7\%).

The failure regions are shared. PartialSpoof-trained detectors reach 23.72--51.38\% EER on the partial-fake sets, and EchoFake's in-domain baselines pass only 65.9--67.6\% of replayed real speech on the open set \cite{partialedit2025,hqmpsd2025,echofake2025}. The difficulties identified here are properties of the conditions, not of a single detector. They also define where future benchmarks are most needed: short partial edits, replay through consumer devices, and real speech from more languages, platforms, and accents.

\subsection{Deploying ARCAS 290M}
The benchmark does not replace validation on a deployment's own audio. Recordings from the actual capture chain (devices, codecs, platforms, and languages) should be used to set operating points, and real-speech scores should be monitored for false-alarm drift. The endpoint's confidence is not a calibrated likelihood ratio, so thresholds should come from validation data. Each decision score should be logged with the integration version and date. The detector output should be one signal in a review process, with human review for high-impact decisions.

\section{Conclusion}
We benchmarked ARCAS 290M, Neural Defend's foundation model for AI-generated audio detection, at full scale on the complete official test sets of ten public benchmarks. On the independent Podonos leaderboard it ranks first of 28 systems, ahead of every other vendor, and on ASVspoof~5 its EER is below that of every commercial system on the Speech DF Arena. On whole-file AI-generated speech, accuracy is 99.32--99.92\% on ASVspoof~5, SpeechFake, and ShiftySpeech, and across five such benchmarks 99.60\% of 4.80 million fakes are detected, including 99.998\% of IndicSynth's Indian-language fakes. Performance holds across codec and telephone conditions, and EER is lower than every zero-shot baseline reported on aligned populations. Condition-level diagnostics place the remaining difficulty in short partial edits, physically replayed audio, already-compressed web audio, and in-the-wild speech in some languages. These conditions remain the hardest for every published detector, so they mark where detection research, benchmark design, and human review should focus next. We recommend that detectors be reported on complete test sets with per-condition breakdowns, not one score per dataset. Comparisons rely on author-reported baselines, PartialEdit and IndicSynth measure recall only, and calibration and adaptive attacks are not assessed.

%% file: sections/09_appendices.tex
\appendix
\renewcommand{\topfraction}{0.95}\renewcommand{\bottomfraction}{0.9}\renewcommand{\textfraction}{0.05}\renewcommand{\floatpagefraction}{0.85}
\raggedbottom
\raggedright
\section{Additional Results}
\label{app:additional_results}

Table~\ref{tab:coverage} lists the release, license, test population, and record accounting for each benchmark.
\input{tables/coverage}

\section{Complete Subset Tables}
\label{app:subset_tables}

The following tables report the slice results summarized in the Results section. Rates are percentages. A dash denotes a quantity that is undefined for a single-class population, and slices with fewer than 30 records per class are omitted or marked.

Table~\ref{tab:shiftyspeech_subsets} reports ShiftySpeech by generator and source domain, and Table~\ref{tab:indicsynth_subsets} reports IndicSynth by tool and language.
\input{tables/shiftyspeech_subsets_inline}
\input{tables/indicsynth_subsets_inline}

Table~\ref{tab:echofake_systems} reports EchoFake recall by generation system.
\input{tables/echofake_subsets_inline}

Table~\ref{tab:ASVspoof5_groups} through Table~\ref{tab:PartialEdit_groups} report every evaluated group in every benchmark: attacks, codecs, generators, replay devices, languages, platforms, and fake-share bands.
\input{tables/group_tables_inline}

\FloatBarrier
\section{Precision--Recall Curves}

\begin{figure}[H]
\centering
\includegraphics[height=0.32\textheight,keepaspectratio]{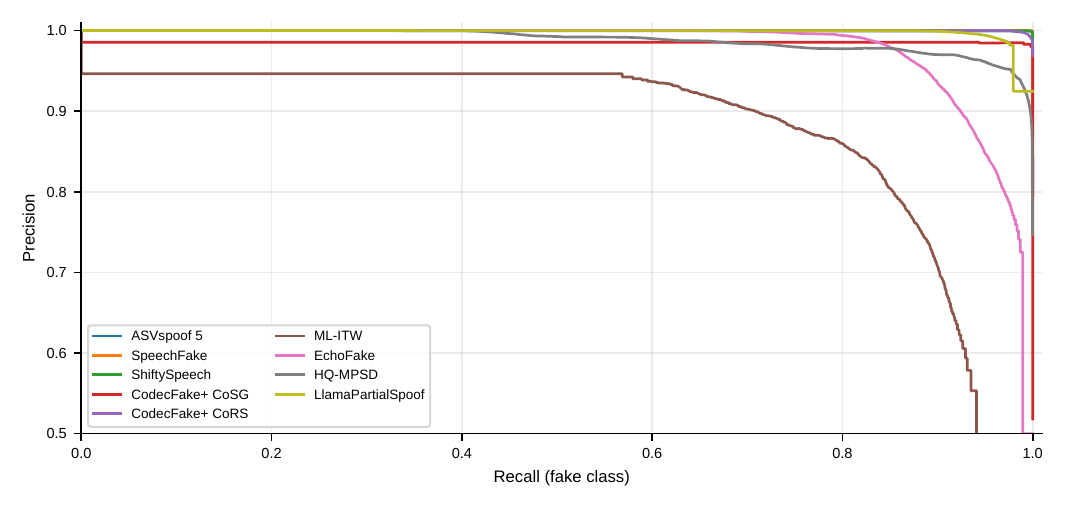}
\caption{Neural Defend ARCAS 290M precision--recall curves across the evaluated benchmark populations cited in the corresponding Results subsections.}
\label{fig:pr_curves}
\end{figure}

\begin{flushleft}
Figure~\ref{fig:pr_curves} compares precision--recall behavior across the benchmarks with real speech. Precision--recall curves are informative when class balance differs across releases, because they retain the trade-off between positive predictive value and fake-speech recall over all score thresholds.
\end{flushleft}

%% file: tables/coverage.tex
\begin{table}[!htbp]
\centering
\caption{Releases, test populations, and record accounting. Unscored records are HTTP errors, undecodable files, or files over 50~MiB; they are not counted as errors.}
\label{tab:coverage}
\footnotesize
\setlength{\tabcolsep}{3.5pt}
\begin{tabular}{@{}llllrrrr@{}}
\toprule
Benchmark & Release used & License & Test population & Scored & Real & Fake & Unscored \\
\midrule
ASVspoof 5 & Zenodo 14498691 & ODC-By 1.0 & Track 1 evaluation & 680,774 & 138,688 & 542,086 & 0 \\
SpeechFake & ModelScope release & CC BY 4.0 & All official test lists & 906,068 & 117,841 & 788,227 & 0 \\
CodecFake+ CoSG & Hugging Face release & MIT & CoSG evaluation & 1,797 & 866 & 931 & 0 \\
CodecFake+ CoRS & Hugging Face release & MIT & CoRS evaluation (p227, p228) & 24,160 & 755 & 23,405 & 0 \\
ML-ITW & Hugging Face release & Not stated & Evaluation split & 24,529 & 18,168 & 6,361 & 0 \\
EchoFake & Hugging Face release & MIT & Closed + open evaluation & 31,591 & 15,798 & 15,793 & 0 \\
HQ-MPSD & Zenodo 17929533 & CC BY 4.0 & All 8 languages & 204,038 & 51,715 & 152,323 & 0 \\
LlamaPartialSpoof & Zenodo 14214149 (v1.0.b) & CC BY 4.0 & Full release, parts a + b & 140,616 & 10,573 & 130,043 & 0 \\
PartialEdit & Zenodo 18829689 (v1.1) & CC BY 4.0 & 70-speaker evaluation split & 107,454 & \NA & 107,454 & 0 \\
ShiftySpeech & Hugging Face release & Apache-2.0 & Released test split (106 archives) & 2,781,458 & 58,806 & 2,722,652 & 0 \\
IndicSynth & Hugging Face release & CC BY-NC 4.0 & Hindi, Tamil, Bengali, Telugu & 741,594 & \NA & 741,594 & 0 \\
\bottomrule
\end{tabular}
\end{table}

%% file: tables/shiftyspeech_subsets_inline.tex
\begin{table}[!htbp]
\centering
\caption{ShiftySpeech: recall (\%) by generator, and recall, specificity, and EER (\%) by source of the real speech. Fake-only sources: Common Voice 99.84 (1,118,664 fakes), train-clean-360 99.64, VoxCeleb 99.64, LJSpeech 99.98, VCTK 99.99.}
\label{tab:shiftyspeech_subsets}
\small
\begin{tabular}{lrr@{\hspace{2.5em}}lrrr}
\toprule
Generator & Fakes & Recall & Source & Recall & Specificity & EER \\
\midrule
TTS, 6 systems & 255,940 & 100.00 & JSUT (Japanese) & 99.98 & 100.00 & 0.02 \\
APNet2 & 63,680 & 95.05 & Podcasts & 99.66 & 99.87 & 0.28 \\
BigVGAN & 201,076 & 97.57 & Audiobooks & 99.49 & 99.87 & 0.38 \\
iSTFTNet & 76,780 & 98.56 & AISHELL (Mandarin) & 98.83 & 99.96 & 0.59 \\
Vocos & 201,076 & 98.79 & YouTube & 93.64 & 95.88 & 5.42 \\
BigVSAN & 201,076 & 98.92 & & & & \\
HiFi-GAN & 304,476 & 99.21 & & & & \\
WaveGrad & 76,780 & 99.56 & & & & \\
UnivNet, MB-MelGAN, & & 99.56 & & & & \\
\quad Style-MelGAN, PWG & & & & & & \\
\bottomrule
\end{tabular}
\end{table}

%% file: tables/indicsynth_subsets_inline.tex
\begin{table}[!htbp]
\centering
\caption{IndicSynth: recall (\%) by generation tool and by language. All 17 misses are XTTS-v2 clips.}
\label{tab:indicsynth_subsets}
\small
\begin{tabular}{lrr@{\hspace{2.5em}}lrr}
\toprule
Tool & Fakes & Recall & Language & Fakes & Recall \\
\midrule
FreeVC24 (VC) & 361,836 & 100.000 & Tamil & 282,312 & 99.999 \\
VITS (Bengali only) & 28,056 & 100.000 & Hindi & 205,938 & 99.998 \\
XTTS-v2 & 351,702 & 99.995 & Telugu & 169,896 & 99.998 \\
 & & & Bengali & 83,448 & 99.993 \\
\bottomrule
\end{tabular}
\end{table}

%% file: tables/echofake_subsets_inline.tex
\begin{table}[!htbp]
\centering
\caption{EchoFake: recall (\%) by generation system, original and replayed fakes pooled (original fakes alone: 99.84\%).}
\label{tab:echofake_systems}
\small
\begin{tabular}{lr@{\hspace{2.5em}}lr}
\toprule
System & Recall & System & Recall \\
\midrule
SpeechT5 & 96.2 & CosyVoice2 & 70.9 \\
OpenVoice-V2 & 94.7 & OpenAudio-S1-mini & 70.7 \\
FireRedTTS & 91.6 & StyleTTS2 & 66.5 \\
XTTS-v2 & 77.6 & MaskGCT & 63.8 \\
IndexTTS & 77.5 & LLaSA & 60.7 \\
F5-TTS & 73.4 & & \\
\bottomrule
\end{tabular}
\end{table}

%% file: tables/group_tables_inline.tex
\begin{center}
\begin{minipage}{\textwidth}
\centering
\captionsetup{type=table,hypcap=false}
\captionof{table}{ASVspoof 5: complete group statistics. Rates are percentages. Mode B compares the group's own real and fake records; mode F compares the group's fakes with all real records of the dataset (TPR, EER, and AUC only). $\dagger$ marks groups with fewer than 30 real or 30 fake records. A dash denotes a quantity undefined for the group.}
\label{tab:ASVspoof5_groups}
\scriptsize
\setlength{\tabcolsep}{3pt}
\begin{tabular*}{\textwidth}{@{\extracolsep{\fill}}lcrrrrrrr}
\toprule
Group & Mode & Fake & Real & Acc. & TPR & TNR & EER & AUC \\
\midrule
attack: A17 & F & 34,414 & 138,688 & -- & 99.90 & -- & 0.15 & 100.00 \\
attack: A18 & F & 34,368 & 138,688 & -- & 100.00 & -- & 0.14 & 100.00 \\
attack: A19 & F & 30,860 & 138,688 & -- & 99.94 & -- & 0.15 & 100.00 \\
attack: A20 & F & 31,444 & 138,688 & -- & 100.00 & -- & 0.13 & 100.00 \\
attack: A21 & F & 34,360 & 138,688 & -- & 99.99 & -- & 0.07 & 100.00 \\
attack: A22 & F & 34,380 & 138,688 & -- & 99.97 & -- & 0.12 & 100.00 \\
attack: A23 & F & 34,392 & 138,688 & -- & 100.00 & -- & 0.07 & 100.00 \\
attack: A24 & F & 34,338 & 138,688 & -- & 99.98 & -- & 0.12 & 100.00 \\
attack: A25 & F & 34,386 & 138,688 & -- & 99.93 & -- & 0.14 & 100.00 \\
attack: A26 & F & 34,386 & 138,688 & -- & 99.93 & -- & 0.14 & 100.00 \\
attack: A27 & F & 34,184 & 138,688 & -- & 99.99 & -- & 0.15 & 100.00 \\
attack: A28 & F & 34,414 & 138,688 & -- & 99.58 & -- & 0.32 & 99.99 \\
attack: A29 & F & 34,408 & 138,688 & -- & 99.99 & -- & 0.06 & 100.00 \\
attack: A30 & F & 34,106 & 138,688 & -- & 100.00 & -- & 0.13 & 100.00 \\
attack: A31 & F & 33,956 & 138,688 & -- & 99.95 & -- & 0.15 & 100.00 \\
attack: A32 & F & 33,690 & 138,688 & -- & 99.99 & -- & 0.13 & 100.00 \\
codec: C00 & B & 136,453 & 35,149 & 99.99 & 100.00 & 99.94 & 0.05 & 100.00 \\
codec: C01 & B & 37,719 & 9,534 & 99.96 & 100.00 & 99.82 & 0.14 & 100.00 \\
codec: C02 & B & 37,772 & 9,654 & 99.93 & 100.00 & 99.67 & 0.17 & 100.00 \\
codec: C03 & B & 38,277 & 9,710 & 99.94 & 99.94 & 99.96 & 0.04 & 100.00 \\
codec: C04 & B & 37,998 & 9,701 & 99.81 & 99.85 & 99.67 & 0.28 & 99.99 \\
codec: C05 & B & 37,848 & 9,693 & 99.99 & 100.00 & 99.94 & 0.03 & 100.00 \\
codec: C06 & B & 32,578 & 8,976 & 99.97 & 100.00 & 99.88 & 0.11 & 100.00 \\
codec: C07 & B & 37,738 & 9,610 & 99.69 & 99.71 & 99.65 & 0.35 & 99.99 \\
codec: C08 & B & 37,675 & 9,003 & 99.79 & 99.81 & 99.70 & 0.29 & 99.99 \\
codec: C09 & B & 32,605 & 8,994 & 99.96 & 99.99 & 99.86 & 0.13 & 100.00 \\
codec: C10 & B & 37,804 & 9,673 & 99.92 & 99.92 & 99.90 & 0.10 & 100.00 \\
codec: C11 & B & 37,619 & 8,991 & 99.97 & 100.00 & 99.86 & 0.10 & 100.00 \\
gender: female & B & 288,042 & 70,236 & 99.92 & 99.95 & 99.81 & 0.17 & 100.00 \\
gender: male & B & 254,044 & 68,452 & 99.92 & 99.94 & 99.88 & 0.12 & 100.00 \\
codec\_q: 0 & B & 136,453 & 35,149 & 99.99 & 100.00 & 99.94 & 0.05 & 100.00 \\
codec\_q: 1 & B & 78,511 & 19,864 & 99.66 & 99.69 & 99.52 & 0.44 & 99.98 \\
codec\_q: 2 & B & 78,408 & 19,931 & 99.93 & 99.95 & 99.82 & 0.14 & 100.00 \\
codec\_q: 3 & B & 78,102 & 20,178 & 99.97 & 99.98 & 99.93 & 0.07 & 100.00 \\
codec\_q: 4 & B & 78,471 & 19,875 & 99.98 & 99.99 & 99.92 & 0.05 & 100.00 \\
codec\_q: 5 & B & 78,075 & 20,297 & 99.96 & 100.00 & 99.83 & 0.14 & 100.00 \\
codec\_q: 6 & B & 4,683 & 1,132 & 99.95 & 100.00 & 99.73 & 0.18 & 100.00 \\
codec\_q: 7 & B & 4,693 & 1,123 & 99.98 & 100.00 & 99.91 & 0.09 & 100.00 \\
codec\_q: 8 & B & 4,690 & 1,139 & 100.00 & 100.00 & 100.00 & 0.00 & 100.00 \\
\bottomrule
\end{tabular*}
\end{minipage}
\end{center}
\medskip

\begin{center}
\begin{minipage}{\textwidth}
\centering
\captionsetup{type=table,hypcap=false}
\captionof{table}{SpeechFake: complete group statistics. Rates are percentages. Mode B compares the group's own real and fake records; mode F compares the group's fakes with all real records of the dataset (TPR, EER, and AUC only). $\dagger$ marks groups with fewer than 30 real or 30 fake records. A dash denotes a quantity undefined for the group.}
\label{tab:SpeechFake_groups}
\scriptsize
\setlength{\tabcolsep}{3pt}
\begin{tabular*}{\textwidth}{@{\extracolsep{\fill}}lcrrrrrrr}
\toprule
Group & Mode & Fake & Real & Acc. & TPR & TNR & EER & AUC \\
\midrule
generator: BigVGAN & F & 12,601 & 117,841 & -- & 94.96 & -- & 2.30 & 99.76 \\
generator: ChatTTS & F & 12,956 & 117,841 & -- & 100.00 & -- & 0.13 & 100.00 \\
generator: CosyVoice & F & 72,087 & 117,841 & -- & 99.99 & -- & 0.10 & 100.00 \\
generator: DiffGANTTS & F & 6,000 & 117,841 & -- & 100.00 & -- & 0.00 & 100.00 \\
generator: EdgeTTS & F & 232,630 & 117,841 & -- & 100.00 & -- & 0.02 & 100.00 \\
generator: FastDiff & F & 12,600 & 117,841 & -- & 100.00 & -- & 0.10 & 100.00 \\
generator: FastSpeech2 & F & 6,000 & 117,841 & -- & 100.00 & -- & 0.00 & 100.00 \\
generator: FireRedTTS & F & 9,781 & 117,841 & -- & 100.00 & -- & 0.00 & 100.00 \\
generator: FishSpeech & F & 9,463 & 117,841 & -- & 100.00 & -- & 0.07 & 100.00 \\
generator: FullBandMelGAN & F & 6,334 & 117,841 & -- & 100.00 & -- & 0.19 & 100.00 \\
generator: GPTSoVITS & F & 10,891 & 117,841 & -- & 100.00 & -- & 0.01 & 100.00 \\
generator: GlowTTS & F & 6,000 & 117,841 & -- & 100.00 & -- & 0.00 & 100.00 \\
generator: GradTTS & F & 5,999 & 117,841 & -- & 100.00 & -- & 0.00 & 100.00 \\
generator: HifiGAN & F & 15,797 & 117,841 & -- & 100.00 & -- & 0.15 & 100.00 \\
generator: MelGAN & F & 7,032 & 117,841 & -- & 100.00 & -- & 0.03 & 100.00 \\
generator: MeloTTS & F & 37,210 & 117,841 & -- & 100.00 & -- & 0.02 & 100.00 \\
generator: OpenVoice & F & 44,366 & 117,841 & -- & 100.00 & -- & 0.01 & 100.00 \\
generator: OpenVoiceTTS & F & 17,747 & 117,841 & -- & 100.00 & -- & 0.00 & 100.00 \\
generator: ParallelWaveGAN & F & 11,284 & 117,841 & -- & 100.00 & -- & 0.09 & 100.00 \\
generator: ParlerTTS & F & 5,417 & 117,841 & -- & 99.98 & -- & 0.09 & 100.00 \\
generator: PortaSpeech & F & 4,564 & 117,841 & -- & 100.00 & -- & 0.00 & 100.00 \\
generator: ProDiffTTS & F & 9,200 & 117,841 & -- & 100.00 & -- & 0.01 & 100.00 \\
generator: SeedVC & F & 161,638 & 117,841 & -- & 100.00 & -- & 0.20 & 100.00 \\
generator: StarGAN & F & 9,642 & 117,841 & -- & 100.00 & -- & 0.00 & 100.00 \\
generator: StyleMelGAN & F & 6,946 & 117,841 & -- & 99.74 & -- & 0.42 & 99.99 \\
generator: StyleTTS2 & F & 5,715 & 117,841 & -- & 100.00 & -- & 0.00 & 100.00 \\
generator: Tacotron2 & F & 6,000 & 117,841 & -- & 100.00 & -- & 0.00 & 100.00 \\
generator: Tortoise & F & 26,248 & 117,841 & -- & 100.00 & -- & 0.04 & 100.00 \\
generator: VITS & F & 6,000 & 117,841 & -- & 100.00 & -- & 0.02 & 100.00 \\
generator: WaveGlow & F & 7,679 & 117,841 & -- & 100.00 & -- & 0.21 & 100.00 \\
generator: WaveNet & F & 2,400 & 117,841 & -- & 100.00 & -- & 0.00 & 100.00 \\
generator\_family: NV & F & 80,273 & 117,841 & -- & 99.19 & -- & 0.64 & 99.96 \\
generator\_family: TTS & F & 434,878 & 117,841 & -- & 100.00 & -- & 0.02 & 100.00 \\
generator\_family: VC & F & 273,076 & 117,841 & -- & 100.00 & -- & 0.18 & 100.00 \\
language: en & B & 225,498 & 41,849 & 99.68 & 99.74 & 99.34 & 0.44 & 99.98 \\
language: es & B & 43,868 & 5,000 & 100.00 & 100.00 & 100.00 & 0.00 & 100.00 \\
language: et & B & 28,392 & 4,958 & 99.99 & 100.00 & 99.96 & 0.02 & 100.00 \\
language: fi & B & 31,619 & 4,959 & 100.00 & 100.00 & 100.00 & 0.00 & 100.00 \\
language: gl & B & 39,838 & 5,000 & 100.00 & 100.00 & 100.00 & 0.00 & 100.00 \\
language: he & B & 20,605 & 605 & 100.00 & 100.00 & 100.00 & 0.00 & 100.00 \\
language: hu & B & 39,981 & 5,000 & 100.00 & 100.00 & 100.00 & 0.00 & 100.00 \\
language: id & B & 39,936 & 4,968 & 100.00 & 100.00 & 100.00 & 0.00 & 100.00 \\
language: is$^\dagger$ & B & 13,373 & 15 & 100.00 & 100.00 & 100.00 & 0.00 & 100.00 \\
language: it & B & 39,989 & 5,000 & 100.00 & 100.00 & 100.00 & 0.00 & 100.00 \\
language: ko & B & 82,728 & 2,500 & 100.00 & 100.00 & 99.96 & 0.00 & 100.00 \\
language: lv & B & 32,845 & 4,939 & 99.98 & 100.00 & 99.84 & 0.10 & 100.00 \\
language: zh & B & 149,555 & 33,048 & 99.85 & 99.95 & 99.38 & 0.35 & 100.00 \\
part: BD & F & 341,767 & 117,841 & -- & 99.81 & -- & 0.34 & 99.99 \\
part: MD & F & 446,460 & 117,841 & -- & 100.00 & -- & 0.13 & 100.00 \\
headline\_set: BD baseline/test\_all & B & 309,065 & 37,242 & 99.70 & 99.79 & 98.96 & 0.56 & 99.98 \\
headline\_set: MD cross\_lingual test & B & 465,291 & 72,944 & 99.99 & 100.00 & 99.95 & 0.03 & 100.00 \\
headline\_set: other experiment test lists only & B & 13,871 & 7,655 & 99.68 & 100.00 & 99.11 & 0.03 & 100.00 \\
\bottomrule
\end{tabular*}
\end{minipage}
\end{center}
\medskip

\begin{center}
\begin{minipage}{\textwidth}
\centering
\captionsetup{type=table,hypcap=false}
\captionof{table}{ShiftySpeech: complete group statistics. Rates are percentages. Mode B compares the group's own real and fake records; mode F compares the group's fakes with all real records of the dataset (TPR, EER, and AUC only). $\dagger$ marks groups with fewer than 30 real or 30 fake records. A dash denotes a quantity undefined for the group.}
\label{tab:ShiftySpeech_groups}
\scriptsize
\setlength{\tabcolsep}{3pt}
\begin{tabular*}{\textwidth}{@{\extracolsep{\fill}}lcrrrrrrr}
\toprule
Group & Mode & Fake & Real & Acc. & TPR & TNR & EER & AUC \\
\midrule
system: FastPitch & F & 4,290 & 58,806 & -- & 100.00 & -- & 0.00 & 100.00 \\
system: Glow-TTS & F & 13,100 & 58,806 & -- & 100.00 & -- & 0.00 & 100.00 \\
system: Grad-TTS & F & 13,100 & 58,806 & -- & 99.97 & -- & 0.30 & 99.99 \\
system: VITS & F & 17,390 & 58,806 & -- & 99.97 & -- & 0.37 & 99.99 \\
system: XTTS & F & 190,670 & 58,806 & -- & 100.00 & -- & 0.05 & 100.00 \\
system: YourTTS & F & 4,290 & 58,806 & -- & 100.00 & -- & 0.00 & 100.00 \\
system: apnet2 & F & 63,680 & 58,806 & -- & 95.05 & -- & 2.85 & 99.65 \\
system: bigvgan & F & 201,076 & 58,806 & -- & 97.57 & -- & 1.81 & 99.80 \\
system: bigvsan & F & 201,076 & 58,806 & -- & 98.92 & -- & 1.17 & 99.93 \\
system: hfg\_vocoded & F & 13,100 & 58,806 & -- & 100.00 & -- & 0.24 & 100.00 \\
system: hifigan & F & 304,476 & 58,806 & -- & 99.21 & -- & 0.94 & 99.95 \\
system: istftnet & F & 76,780 & 58,806 & -- & 98.56 & -- & 1.36 & 99.93 \\
system: mb-melgan & F & 304,476 & 58,806 & -- & 99.84 & -- & 0.45 & 99.99 \\
system: pwg & F & 317,564 & 58,806 & -- & 99.96 & -- & 0.24 & 100.00 \\
system: style-melgan & F & 317,576 & 58,806 & -- & 99.90 & -- & 0.38 & 99.99 \\
system: univnet-c16 & F & 201,076 & 58,806 & -- & 99.80 & -- & 0.59 & 99.98 \\
system: univnet-c32 & F & 201,076 & 58,806 & -- & 99.83 & -- & 0.59 & 99.98 \\
system: vocos & F & 201,076 & 58,806 & -- & 98.79 & -- & 1.20 & 99.91 \\
system: wavegrad & F & 76,780 & 58,806 & -- & 99.56 & -- & 0.74 & 99.97 \\
generator\_family: TTS & F & 255,940 & 58,806 & -- & 100.00 & -- & 0.10 & 100.00 \\
generator\_family: vocoder & F & 2,466,712 & 58,806 & -- & 99.26 & -- & 0.98 & 99.95 \\
source\_domain: aishell & B & 86,112 & 7,176 & 98.92 & 98.83 & 99.96 & 0.59 & 99.93 \\
source\_domain: aishell-dev & B & 14,326 & 58,806 & -- & 100.00 & -- & 0.06 & 100.00 \\
source\_domain: aishell-test & B & 7,176 & 58,806 & -- & 100.00 & -- & 0.06 & 100.00 \\
source\_domain: aishell-train & B & 120,098 & 58,806 & -- & 100.00 & -- & 0.06 & 100.00 \\
source\_domain: audiobook & B & 133,500 & 11,126 & 99.52 & 99.49 & 99.87 & 0.38 & 99.98 \\
source\_domain: commonvoice & B & 1,118,664 & 58,806 & -- & 99.84 & -- & 0.63 & 99.97 \\
source\_domain: jsut & B & 65,000 & 5,000 & 99.98 & 99.98 & 100.00 & 0.02 & 100.00 \\
source\_domain: ljspeech & B & 170,300 & 58,806 & -- & 99.98 & -- & 0.21 & 100.00 \\
source\_domain: podcast & B & 237,780 & 19,815 & 99.68 & 99.66 & 99.87 & 0.28 & 99.99 \\
source\_domain: train-clean-360 & B & 466,000 & 58,806 & -- & 99.64 & -- & 0.50 & 99.98 \\
source\_domain: vctk & B & 56,940 & 58,806 & -- & 99.99 & -- & 0.01 & 100.00 \\
source\_domain: voxceleb\_test & B & 58,488 & 58,806 & -- & 99.64 & -- & 0.98 & 99.97 \\
source\_domain: youtube & B & 188,268 & 15,689 & 93.81 & 93.64 & 95.88 & 5.42 & 99.01 \\
language: en & B & 2,429,940 & 46,630 & 99.28 & 99.29 & 98.53 & 1.05 & 99.94 \\
language: ja & B & 65,000 & 5,000 & 99.98 & 99.98 & 100.00 & 0.02 & 100.00 \\
language: zh & B & 227,712 & 7,176 & 99.57 & 99.56 & 99.96 & 0.25 & 99.97 \\
\bottomrule
\end{tabular*}
\end{minipage}
\end{center}
\medskip

\begin{center}
\begin{minipage}{\textwidth}
\centering
\captionsetup{type=table,hypcap=false}
\captionof{table}{CodecFake+ CoSG: complete group statistics. Rates are percentages. Mode B compares the group's own real and fake records; mode F compares the group's fakes with all real records of the dataset (TPR, EER, and AUC only). $\dagger$ marks groups with fewer than 30 real or 30 fake records. A dash denotes a quantity undefined for the group.}
\label{tab:CodecFakeCoSG_groups}
\scriptsize
\setlength{\tabcolsep}{3pt}
\begin{tabular*}{\textwidth}{@{\extracolsep{\fill}}lcrrrrrrr}
\toprule
Group & Mode & Fake & Real & Acc. & TPR & TNR & EER & AUC \\
\midrule
system: CLAMTTS & B & 78 & 41 & 97.48 & 100.00 & 92.68 & 2.38 & 98.78 \\
system: ELLAV$^\dagger$ & B & 8 & 8 & 100.00 & 100.00 & 100.00 & 0.00 & 100.00 \\
system: GPST$^\dagger$ & B & 30 & 10 & 95.00 & 93.33 & 100.00 & 0.00 & 100.00 \\
system: MASKGCT & B & 576 & 576 & 99.91 & 100.00 & 99.83 & 0.00 & 100.00 \\
system: NS2$^\dagger$ & B & 23 & 23 & 100.00 & 100.00 & 100.00 & 0.00 & 100.00 \\
system: NS3 & B & 32 & 32 & 100.00 & 100.00 & 100.00 & 0.00 & 100.00 \\
system: RALLE$^\dagger$ & B & 5 & 5 & 100.00 & 100.00 & 100.00 & 0.00 & 100.00 \\
system: SIMPLESPEECH1 & B & 32 & 34 & 96.97 & 100.00 & 94.12 & 5.56 & 97.06 \\
system: SIMPLESPEECH2 & B & 31 & 31 & 98.39 & 100.00 & 96.77 & 3.12 & 98.39 \\
system: SINGLECODEC$^\dagger$ & B & 14 & 8 & 100.00 & 100.00 & 100.00 & 0.00 & 100.00 \\
system: SPEECHX$^\dagger$ & B & 8 & 8 & 100.00 & 100.00 & 100.00 & 0.00 & 100.00 \\
system: TACOLM$^\dagger$ & B & 8 & 8 & 100.00 & 100.00 & 100.00 & 0.00 & 100.00 \\
system: TI1G1R$^\dagger$ & B & 6 & 6 & 100.00 & 100.00 & 100.00 & 0.00 & 100.00 \\
system: UNIAUDIO$^\dagger$ & B & 11 & 11 & 59.09 & 100.00 & 18.18 & 45.00 & 59.09 \\
system: USLM$^\dagger$ & B & 5 & 5 & 100.00 & 100.00 & 100.00 & 0.00 & 100.00 \\
system: VALLE & B & 57 & 53 & 100.00 & 100.00 & 100.00 & 0.00 & 100.00 \\
system: VIOLA$^\dagger$ & B & 7 & 7 & 100.00 & 100.00 & 100.00 & 0.00 & 100.00 \\
quantiser: Mvq & F & 848 & 866 & -- & 99.76 & -- & 1.63 & 99.19 \\
quantiser: Scq & F & 63 & 866 & -- & 100.00 & -- & 1.48 & 99.25 \\
quantiser: Svq$^\dagger$ & F & 20 & 866 & -- & 100.00 & -- & 1.62 & 99.21 \\
aux\_objective: Disent & F & 52 & 866 & -- & 100.00 & -- & 1.77 & 99.18 \\
aux\_objective: nan & F & 874 & 866 & -- & 99.77 & -- & 1.62 & 99.20 \\
aux\_objective: Sem$^\dagger$ & F & 5 & 866 & -- & 100.00 & -- & 1.62 & 99.08 \\
decoder\_domain: Freq & F & 668 & 866 & -- & 100.00 & -- & 1.56 & 99.23 \\
decoder\_domain: Time & F & 263 & 866 & -- & 99.24 & -- & 1.87 & 99.10 \\
\bottomrule
\end{tabular*}
\end{minipage}
\end{center}
\medskip

\begin{center}
\begin{minipage}{\textwidth}
\centering
\captionsetup{type=table,hypcap=false}
\captionof{table}{CodecFake+ CoRS: complete group statistics. Rates are percentages. Mode B compares the group's own real and fake records; mode F compares the group's fakes with all real records of the dataset (TPR, EER, and AUC only). $\dagger$ marks groups with fewer than 30 real or 30 fake records. A dash denotes a quantity undefined for the group.}
\label{tab:CodecFakeCoRS_groups}
\scriptsize
\setlength{\tabcolsep}{3pt}
\begin{tabular*}{\textwidth}{@{\extracolsep{\fill}}lcrrrrrrr}
\toprule
Group & Mode & Fake & Real & Acc. & TPR & TNR & EER & AUC \\
\midrule
codec: DAC24 & F & 755 & 755 & -- & 66.49 & -- & 13.64 & 94.14 \\
codec: Encodec\_24b24k & F & 755 & 755 & -- & 100.00 & -- & 1.06 & 99.89 \\
codec: Encodec\_6b24k & F & 755 & 755 & -- & 100.00 & -- & 0.26 & 99.93 \\
codec: FACodec\_encodec-decoder-v2\_16k & F & 755 & 755 & -- & 100.00 & -- & 0.79 & 99.91 \\
codec: FunCodec\_en-libritts-16k-gr1nq32ds320 & F & 755 & 755 & -- & 100.00 & -- & 0.13 & 99.93 \\
codec: FunCodec\_zh\_en-general-16k-nq32ds320 & F & 755 & 755 & -- & 100.00 & -- & 0.13 & 99.93 \\
codec: HiFi\_Codec\_16k\_320d\_large\_universal & F & 755 & 755 & -- & 100.00 & -- & 0.13 & 99.93 \\
codec: SemantiCodec\_0.35kbps\_16k & F & 755 & 755 & -- & 100.00 & -- & 0.26 & 99.93 \\
codec: SemantiCodec\_0.70kbps\_16k & F & 755 & 755 & -- & 100.00 & -- & 0.26 & 99.93 \\
codec: SemantiCodec\_1.40kbps\_16k & F & 755 & 755 & -- & 100.00 & -- & 0.26 & 99.93 \\
codec: SpeechTokenizer\_hubert\_avg & F & 755 & 755 & -- & 100.00 & -- & 0.26 & 99.93 \\
codec: audiodec\_24k\_320d & F & 755 & 755 & -- & 100.00 & -- & 0.13 & 99.93 \\
codec: bigcodec & F & 755 & 755 & -- & 99.74 & -- & 1.99 & 99.80 \\
codec: hilcodec\_speech & F & 755 & 755 & -- & 100.00 & -- & 0.40 & 99.93 \\
codec: languagecodec & F & 755 & 755 & -- & 100.00 & -- & 0.53 & 99.91 \\
codec: llm\_codec & F & 755 & 755 & -- & 100.00 & -- & 0.13 & 99.93 \\
codec: mini & F & 755 & 755 & -- & 100.00 & -- & 0.26 & 99.93 \\
codec: snac\_24khz & F & 755 & 755 & -- & 100.00 & -- & 0.26 & 99.93 \\
codec: snac\_44khz & F & 755 & 755 & -- & 100.00 & -- & 0.13 & 99.93 \\
codec: socodec\_16384x4\_120ms\_16khz\_chinese & F & 755 & 755 & -- & 100.00 & -- & 0.13 & 99.93 \\
codec: spectralcodecs & F & 755 & 755 & -- & 100.00 & -- & 0.40 & 99.92 \\
codec: sqcodec50dim9 & F & 755 & 755 & -- & 91.26 & -- & 6.36 & 98.28 \\
codec: ticodec\_1g1r & F & 755 & 755 & -- & 100.00 & -- & 0.13 & 99.93 \\
codec: ticodec\_1g2r & F & 755 & 755 & -- & 100.00 & -- & 0.13 & 99.93 \\
codec: ticodec\_1g4r & F & 755 & 755 & -- & 100.00 & -- & 0.13 & 99.93 \\
codec: vocos\_encodec\_12 & F & 755 & 755 & -- & 100.00 & -- & 0.53 & 99.92 \\
codec: vocos\_encodec\_6 & F & 755 & 755 & -- & 100.00 & -- & 0.26 & 99.93 \\
codec: wavtokenizer\_medium\_320\_24k\_4096 & F & 755 & 755 & -- & 100.00 & -- & 0.26 & 99.93 \\
codec: wavtokenizer\_small\_320\_24k\_4096 & F & 755 & 755 & -- & 100.00 & -- & 0.26 & 99.93 \\
codec: xcodec\_hubert\_libirispeech & F & 755 & 755 & -- & 100.00 & -- & 0.13 & 99.93 \\
codec: xocdec\_hubert\_general\_audio & F & 755 & 755 & -- & 100.00 & -- & 0.13 & 99.93 \\
\bottomrule
\end{tabular*}
\end{minipage}
\end{center}
\medskip

\begin{center}
\begin{minipage}{\textwidth}
\centering
\captionsetup{type=table,hypcap=false}
\captionof{table}{ML-ITW: complete group statistics. Rates are percentages. Mode B compares the group's own real and fake records; mode F compares the group's fakes with all real records of the dataset (TPR, EER, and AUC only). $\dagger$ marks groups with fewer than 30 real or 30 fake records. A dash denotes a quantity undefined for the group.}
\label{tab:MLITW_groups}
\scriptsize
\setlength{\tabcolsep}{3pt}
\begin{tabular*}{\textwidth}{@{\extracolsep{\fill}}lcrrrrrrr}
\toprule
Group & Mode & Fake & Real & Acc. & TPR & TNR & EER & AUC \\
\midrule
language: chinese & B & 3,074 & 5,399 & 86.52 & 78.79 & 90.92 & 12.26 & 93.27 \\
language: english & B & 1,598 & 6,931 & 94.72 & 82.48 & 97.55 & 9.14 & 95.32 \\
language: french & B & 61 & 195 & 92.97 & 85.25 & 95.38 & 14.75 & 90.90 \\
language: german & B & 356 & 704 & 89.06 & 76.69 & 95.31 & 18.58 & 88.30 \\
language: hebrew$^\dagger$ & B & 13 & 36 & 93.88 & 76.92 & 100.00 & 0.00 & 100.00 \\
language: hindi & B & 358 & 292 & 88.31 & 79.05 & 99.66 & 15.17 & 91.03 \\
language: hungarian$^\dagger$ & B & 3 & 296 & 96.66 & 33.33 & 97.30 & 4.01 & 97.69 \\
language: japanese & B & 277 & 1,292 & 95.92 & 81.95 & 98.92 & 7.21 & 98.22 \\
language: korean & B & 110 & 818 & 96.12 & 69.09 & 99.76 & 15.30 & 90.42 \\
language: portuguese & B & 108 & 911 & 93.92 & 56.48 & 98.35 & 28.66 & 79.74 \\
language: russian & B & 44 & 329 & 90.88 & 22.73 & 100.00 & 28.49 & 79.50 \\
language: spanish & B & 166 & 480 & 98.30 & 96.99 & 98.75 & 2.41 & 99.10 \\
language: turkish & B & 176 & 412 & 83.67 & 60.23 & 93.69 & 26.14 & 80.99 \\
language: ukrainian$^\dagger$ & B & 17 & 73 & 97.78 & 88.24 & 100.00 & 0.00 & 100.00 \\
platform: bilibili & B & 793 & 4,263 & 87.92 & 91.80 & 87.19 & 11.46 & 94.48 \\
platform: douyin & B & 1,149 & 182 & 80.47 & 77.46 & 99.45 & 8.25 & 97.26 \\
platform: facebook & B & 1,129 & 3,027 & 92.11 & 78.83 & 97.06 & 15.46 & 90.31 \\
platform: rednote & B & 1,393 & 1,333 & 81.88 & 66.33 & 98.12 & 14.55 & 93.16 \\
platform: tiktok & B & 305 & 369 & 96.44 & 92.79 & 99.46 & 5.25 & 97.46 \\
platform: x & B & 1,061 & 3,046 & 95.13 & 85.39 & 98.52 & 10.93 & 94.25 \\
platform: youtube & B & 531 & 5,948 & 96.87 & 74.20 & 98.89 & 8.17 & 95.65 \\
\bottomrule
\end{tabular*}
\end{minipage}
\end{center}
\medskip

\begin{center}
\begin{minipage}{\textwidth}
\centering
\captionsetup{type=table,hypcap=false}
\captionof{table}{EchoFake: complete group statistics. Rates are percentages. Mode B compares the group's own real and fake records; mode F compares the group's fakes with all real records of the dataset (TPR, EER, and AUC only). $\dagger$ marks groups with fewer than 30 real or 30 fake records. A dash denotes a quantity undefined for the group.}
\label{tab:EchoFake_groups}
\scriptsize
\setlength{\tabcolsep}{3pt}
\begin{tabular*}{\textwidth}{@{\extracolsep{\fill}}lcrrrrrrr}
\toprule
Group & Mode & Fake & Real & Acc. & TPR & TNR & EER & AUC \\
\midrule
label\_4class: bonafide & B & 0 & 7,900 & 99.94 & -- & 99.94 & -- & -- \\
label\_4class: fake & B & 7,900 & 15,798 & -- & 99.84 & -- & 0.24 & 99.99 \\
label\_4class: replay\_bonafide & B & 0 & 7,898 & 99.37 & -- & 99.37 & -- & -- \\
label\_4class: replay\_fake & B & 7,893 & 15,798 & -- & 57.62 & -- & 12.79 & 95.03 \\
eval\_set: closed & B & 2,993 & 2,998 & 86.71 & 74.37 & 99.03 & 10.83 & 96.24 \\
eval\_set: open & B & 12,800 & 12,800 & 89.78 & 79.76 & 99.80 & 7.74 & 97.79 \\
generator: CosyVoice2 & F & 2,567 & 15,798 & -- & 70.90 & -- & 11.32 & 95.73 \\
generator: F5-TTS & F & 478 & 15,798 & -- & 73.43 & -- & 6.90 & 98.11 \\
generator: FireRedTTS & F & 2,548 & 15,798 & -- & 91.56 & -- & 3.65 & 99.38 \\
generator: IndexTTS & F & 2,536 & 15,798 & -- & 77.52 & -- & 9.01 & 97.14 \\
generator: LLaSA & F & 501 & 15,798 & -- & 60.68 & -- & 10.43 & 96.74 \\
generator: MaskGCT & F & 2,546 & 15,798 & -- & 63.83 & -- & 12.74 & 95.31 \\
generator: OpenAudio-S1-mini & F & 468 & 15,798 & -- & 70.73 & -- & 10.52 & 96.85 \\
generator: OpenVoice-V2 & F & 2,603 & 15,798 & -- & 94.70 & -- & 3.76 & 99.35 \\
generator: SpeechT5 & F & 519 & 15,798 & -- & 96.15 & -- & 1.35 & 99.89 \\
generator: StyleTTS2 & F & 505 & 15,798 & -- & 66.53 & -- & 9.47 & 97.43 \\
generator: XTTS-v2 & F & 522 & 15,798 & -- & 77.59 & -- & 5.56 & 98.96 \\
replay\_room: 12.8m(L) $^*$ 9.3m(W) $^*$ 3.2m(H) & B & 747 & 745 & 73.26 & 49.26 & 97.32 & 22.82 & 85.43 \\
replay\_room: 18.6m(L) $^*$ 13.2m(W) $^*$ 3.2m(H) & B & 6,400 & 6,400 & 79.66 & 59.64 & 99.67 & 14.30 & 93.64 \\
replay\_room: 4.8m(L) $^*$ 3.2m(W) $^*$ 3.2m(H) & B & 746 & 753 & 73.85 & 48.66 & 98.80 & 20.88 & 87.13 \\
replay\_room: none & B & 7,900 & 7,900 & 99.89 & 99.84 & 99.94 & 0.15 & 99.99 \\
replay\_player: Edifier MR4 Powered Studio Monitor Speakers & B & 3,200 & 3,200 & 81.36 & 62.91 & 99.81 & 10.25 & 95.70 \\
replay\_player: MacBook Pro 14-inch 2021 & B & 748 & 752 & 74.40 & 50.00 & 98.67 & 19.95 & 88.38 \\
replay\_player: Xiaomi 13 Ultra & B & 3,200 & 3,200 & 77.95 & 56.38 & 99.53 & 17.22 & 91.46 \\
replay\_player: iPad Mini (7th generation) & B & 745 & 746 & 72.70 & 47.92 & 97.45 & 22.67 & 84.48 \\
replay\_player: none & B & 7,900 & 7,900 & 99.89 & 99.84 & 99.94 & 0.15 & 99.99 \\
replay\_recorder: MacBook Pro 14-inch 2021 & B & 3,200 & 3,200 & 84.16 & 68.59 & 99.72 & 11.28 & 95.63 \\
replay\_recorder: Samsung Galaxy A54 & B & 745 & 753 & 71.23 & 43.76 & 98.41 & 24.10 & 83.60 \\
replay\_recorder: Wired Earbuds With Microphone & B & 1,600 & 1,600 & 68.66 & 38.06 & 99.25 & 25.61 & 82.90 \\
replay\_recorder: Xiaomi 13 Ultra & B & 1,600 & 1,600 & 81.66 & 63.31 & 100.00 & 7.03 & 96.89 \\
replay\_recorder: iPhone 13 mini & B & 748 & 745 & 75.89 & 54.14 & 97.72 & 18.29 & 89.25 \\
replay\_recorder: none & B & 7,900 & 7,900 & 99.89 & 99.84 & 99.94 & 0.15 & 99.99 \\
replay\_distance: 15cm & B & 747 & 745 & 76.34 & 53.95 & 98.79 & 16.78 & 90.58 \\
replay\_distance: 30cm & B & 6,400 & 6,400 & 79.66 & 59.64 & 99.67 & 14.30 & 93.64 \\
replay\_distance: 50cm & B & 746 & 753 & 70.78 & 43.97 & 97.34 & 25.68 & 81.26 \\
replay\_distance: none & B & 7,900 & 7,900 & 99.89 & 99.84 & 99.94 & 0.15 & 99.99 \\
fake\_type: full & B & 7,900 & 15,798 & -- & 99.84 & -- & 0.24 & 99.99 \\
fake\_type: real & B & 0 & 7,900 & 99.94 & -- & 99.94 & -- & -- \\
fake\_type: replayed & B & 7,893 & 7,898 & 78.50 & 57.62 & 99.37 & 16.29 & 92.29 \\
\bottomrule
\end{tabular*}
\end{minipage}
\end{center}
\medskip

\begin{center}
\begin{minipage}{\textwidth}
\centering
\captionsetup{type=table,hypcap=false}
\captionof{table}{HQ-MPSD: complete group statistics. Rates are percentages. Mode B compares the group's own real and fake records; mode F compares the group's fakes with all real records of the dataset (TPR, EER, and AUC only). $\dagger$ marks groups with fewer than 30 real or 30 fake records. A dash denotes a quantity undefined for the group.}
\label{tab:HQMPSD_groups}
\scriptsize
\setlength{\tabcolsep}{3pt}
\begin{tabular*}{\textwidth}{@{\extracolsep{\fill}}lcrrrrrrr}
\toprule
Group & Mode & Fake & Real & Acc. & TPR & TNR & EER & AUC \\
\midrule
language: Dutch & B & 3,840 & 1,280 & 84.75 & 79.66 & 100.00 & 2.81 & 99.35 \\
language: English & B & 15,487 & 6,103 & 94.84 & 92.96 & 99.59 & 1.74 & 99.83 \\
language: French & B & 26,511 & 8,837 & 85.72 & 81.42 & 98.62 & 5.19 & 98.82 \\
language: German & B & 24,723 & 8,241 & 84.44 & 79.34 & 99.75 & 2.45 & 99.63 \\
language: Italian & B & 12,036 & 4,012 & 84.42 & 79.74 & 98.48 & 4.71 & 98.88 \\
language: Polish & B & 29,607 & 9,869 & 88.26 & 91.79 & 77.68 & 14.77 & 93.54 \\
language: Portuguese & B & 12,000 & 4,000 & 77.21 & 69.62 & 99.98 & 3.15 & 99.65 \\
language: Spanish & B & 28,119 & 9,373 & 86.37 & 83.31 & 95.56 & 7.10 & 97.42 \\
subset: Fully\_Fake & F & 51,715 & 51,715 & -- & 100.00 & -- & 0.08 & 100.00 \\
subset: Partial\_Fake\_Clean & F & 51,715 & 51,715 & -- & 83.15 & -- & 8.36 & 96.67 \\
subset: Partial\_Fake\_Noisy & F & 48,893 & 51,715 & -- & 66.46 & -- & 11.51 & 94.56 \\
fake\_type: full & F & 51,715 & 51,715 & -- & 100.00 & -- & 0.08 & 100.00 \\
fake\_type: partial & F & 100,608 & 51,715 & -- & 75.04 & -- & 9.93 & 95.64 \\
condition: Background Noise & F & 15,721 & 51,715 & -- & 78.33 & -- & 8.92 & 96.15 \\
condition: RIR & F & 14,934 & 51,715 & -- & 61.06 & -- & 12.29 & 93.96 \\
condition: RIR+Background Noise & F & 14,957 & 51,715 & -- & 56.33 & -- & 13.13 & 93.14 \\
condition: clean & F & 103,430 & 51,715 & -- & 91.58 & -- & 6.34 & 98.33 \\
condition: noisy (type not documented) & F & 3,281 & 51,715 & -- & 80.31 & -- & 9.14 & 96.12 \\
fake\_share\_bucket: 10-25\% & F & 704 & 51,715 & -- & 20.45 & -- & 14.91 & 88.99 \\
fake\_share\_bucket: 25-50\% & F & 79,340 & 51,715 & -- & 72.76 & -- & 10.10 & 95.38 \\
fake\_share\_bucket: <10\% & F & 54 & 51,715 & -- & 1.85 & -- & 37.04 & 63.76 \\
fake\_share\_bucket: >=50\% & F & 11,126 & 51,715 & -- & 83.81 & -- & 9.51 & 96.69 \\
fake\_share\_bucket: unknown & F & 9,384 & 51,715 & -- & 88.38 & -- & 7.55 & 97.29 \\
fake\_share\_bucket: whole clip & F & 51,715 & 51,715 & -- & 100.00 & -- & 0.08 & 100.00 \\
\bottomrule
\end{tabular*}
\end{minipage}
\end{center}
\medskip

\begin{center}
\begin{minipage}{\textwidth}
\centering
\captionsetup{type=table,hypcap=false}
\captionof{table}{LlamaPartialSpoof: complete group statistics. Rates are percentages. Mode B compares the group's own real and fake records; mode F compares the group's fakes with all real records of the dataset (TPR, EER, and AUC only). $\dagger$ marks groups with fewer than 30 real or 30 fake records. A dash denotes a quantity undefined for the group.}
\label{tab:LlamaPartialSpoof_groups}
\scriptsize
\setlength{\tabcolsep}{3pt}
\begin{tabular*}{\textwidth}{@{\extracolsep{\fill}}lcrrrrrrr}
\toprule
Group & Mode & Fake & Real & Acc. & TPR & TNR & EER & AUC \\
\midrule
fake\_type: full & F & 33,461 & 10,573 & -- & 99.97 & -- & 0.02 & 100.00 \\
fake\_type: partial & F & 96,582 & 10,573 & -- & 75.05 & -- & 6.80 & 97.56 \\
tool: cosyvoice & F & 21,362 & 10,573 & -- & 87.39 & -- & 3.86 & 98.92 \\
tool: elevenlab & F & 21,738 & 10,573 & -- & 79.69 & -- & 5.62 & 98.25 \\
tool: gptsovits & F & 21,738 & 10,573 & -- & 82.78 & -- & 5.59 & 98.16 \\
tool: ljjets & F & 21,738 & 10,573 & -- & 66.34 & -- & 9.08 & 96.30 \\
tool: xttsv2 & F & 21,729 & 10,573 & -- & 84.09 & -- & 4.83 & 98.61 \\
tool: yourtts & F & 21,738 & 10,573 & -- & 88.58 & -- & 3.82 & 98.92 \\
joining\_method: crossfade & F & 32,194 & 10,573 & -- & 75.90 & -- & 6.61 & 97.67 \\
joining\_method: cut/paste & F & 32,194 & 10,573 & -- & 77.33 & -- & 5.67 & 98.26 \\
joining\_method: n/a (full) & F & 33,461 & 10,573 & -- & 99.97 & -- & 0.02 & 100.00 \\
joining\_method: overlap/add & F & 32,194 & 10,573 & -- & 71.92 & -- & 8.11 & 96.75 \\
crossfade\_function: half-sine & F & 6,390 & 10,573 & -- & 76.28 & -- & 6.42 & 97.81 \\
crossfade\_function: inverted-parabola & F & 6,467 & 10,573 & -- & 75.48 & -- & 6.43 & 97.81 \\
crossfade\_function: logarithmic & F & 6,472 & 10,573 & -- & 75.46 & -- & 6.94 & 97.58 \\
crossfade\_function: none & F & 97,849 & 10,573 & -- & 83.29 & -- & 5.26 & 98.36 \\
crossfade\_function: quarter-sine & F & 6,515 & 10,573 & -- & 76.18 & -- & 6.54 & 97.71 \\
crossfade\_function: triangle & F & 6,350 & 10,573 & -- & 76.09 & -- & 6.83 & 97.47 \\
fake\_share\_bucket: 10-25\% & F & 10,304 & 10,573 & -- & 26.73 & -- & 12.89 & 93.05 \\
fake\_share\_bucket: 25-50\% & F & 24,761 & 10,573 & -- & 65.17 & -- & 6.62 & 97.92 \\
fake\_share\_bucket: <10\% & F & 5,378 & 10,573 & -- & 4.00 & -- & 22.61 & 81.96 \\
fake\_share\_bucket: >=50\% & F & 56,139 & 10,573 & -- & 95.08 & -- & 1.75 & 99.73 \\
fake\_share\_bucket: whole clip & F & 33,461 & 10,573 & -- & 99.97 & -- & 0.02 & 100.00 \\
part: R01TTS.0.a & F & 65,655 & 10,573 & -- & 88.16 & -- & 4.20 & 98.86 \\
part: R01TTS.0.b & F & 64,388 & 10,573 & -- & 74.63 & -- & 6.90 & 97.51 \\
\bottomrule
\end{tabular*}
\end{minipage}
\end{center}
\medskip

\begin{center}
\begin{minipage}{\textwidth}
\centering
\captionsetup{type=table,hypcap=false}
\captionof{table}{IndicSynth: complete group statistics. Rates are percentages. Mode B compares the group's own real and fake records; mode F compares the group's fakes with all real records of the dataset (TPR, EER, and AUC only). $\dagger$ marks groups with fewer than 30 real or 30 fake records. A dash denotes a quantity undefined for the group.}
\label{tab:IndicSynth_groups}
\scriptsize
\setlength{\tabcolsep}{3pt}
\begin{tabular*}{\textwidth}{@{\extracolsep{\fill}}lcrrrrrrr}
\toprule
Group & Mode & Fake & Real & Acc. & TPR & TNR & EER & AUC \\
\midrule
language: Bengali & F & 83,448 & 0 & -- & 99.99 & -- & -- & -- \\
language: Hindi & F & 205,938 & 0 & -- & 100.00 & -- & -- & -- \\
language: Tamil & F & 282,312 & 0 & -- & 100.00 & -- & -- & -- \\
language: Telugu & F & 169,896 & 0 & -- & 100.00 & -- & -- & -- \\
generator: freevc24 & F & 361,836 & 0 & -- & 100.00 & -- & -- & -- \\
generator: vits & F & 28,056 & 0 & -- & 100.00 & -- & -- & -- \\
generator: xtts\_v2 & F & 351,702 & 0 & -- & 100.00 & -- & -- & -- \\
gender: Female & F & 310,874 & 0 & -- & 100.00 & -- & -- & -- \\
gender: Male & F & 430,720 & 0 & -- & 100.00 & -- & -- & -- \\
subset: diversity-proxy (TTS: xtts\_v2/vits) & F & 379,758 & 0 & -- & 100.00 & -- & -- & -- \\
subset: mimicry-proxy (VC: freevc24) & F & 361,836 & 0 & -- & 100.00 & -- & -- & -- \\
\bottomrule
\end{tabular*}
\end{minipage}
\end{center}
\medskip

\begin{center}
\begin{minipage}{\textwidth}
\centering
\captionsetup{type=table,hypcap=false}
\captionof{table}{PartialEdit: complete group statistics. Rates are percentages. Mode B compares the group's own real and fake records; mode F compares the group's fakes with all real records of the dataset (TPR, EER, and AUC only). $\dagger$ marks groups with fewer than 30 real or 30 fake records. A dash denotes a quantity undefined for the group.}
\label{tab:PartialEdit_groups}
\scriptsize
\setlength{\tabcolsep}{3pt}
\begin{tabular*}{\textwidth}{@{\extracolsep{\fill}}lcrrrrrrr}
\toprule
Group & Mode & Fake & Real & Acc. & TPR & TNR & EER & AUC \\
\midrule
editor: E1 & F & 53,682 & 0 & -- & 65.52 & -- & -- & -- \\
editor: E2 & F & 53,772 & 0 & -- & 68.29 & -- & -- & -- \\
codec: codec & F & 53,727 & 0 & -- & 100.00 & -- & -- & -- \\
codec: no codec & F & 53,727 & 0 & -- & 33.81 & -- & -- & -- \\
fake\_share\_bucket: 10-25\% & F & 37,210 & 0 & -- & 24.61 & -- & -- & -- \\
fake\_share\_bucket: 25-50\% & F & 13,595 & 0 & -- & 62.62 & -- & -- & -- \\
fake\_share\_bucket: <10\% & F & 2,451 & 0 & -- & 2.41 & -- & -- & -- \\
fake\_share\_bucket: >=50\% & F & 471 & 0 & -- & 92.78 & -- & -- & -- \\
fake\_share\_bucket: whole clip (codec) & F & 53,727 & 0 & -- & 100.00 & -- & -- & -- \\
subset: E1 & F & 26,841 & 0 & -- & 31.04 & -- & -- & -- \\
subset: E1-Codec & F & 26,841 & 0 & -- & 100.00 & -- & -- & -- \\
subset: E2 & F & 26,886 & 0 & -- & 36.58 & -- & -- & -- \\
subset: E2-Codec & F & 26,886 & 0 & -- & 100.00 & -- & -- & -- \\
\bottomrule
\end{tabular*}
\end{minipage}
\end{center}
\medskip